\documentclass[a4paper,11pt]{article}
\usepackage[english]{babel}
\RequirePackage{snapshot}  

\usepackage{bm,amsmath,amssymb,slashed,graphicx,%
            enumerate,alltt,xspace,multirow,xcolor,mathrsfs}
\usepackage{fancyvrb}
\usepackage{booktabs,tabularx}
\usepackage{graphicx}
\usepackage{subcaption}
\usepackage{xspace}
\usepackage{cancel}
\usepackage[utf8]{inputenc} 
\usepackage{scalerel}
\usepackage{layouts}
\usepackage{adjustbox}
\usepackage{pdflscape}
\usepackage[normalem]{ulem} 
\usepackage{lineno}
\usepackage{wasysym}

\usepackage{algorithm}
\usepackage[noEnd=true]{algpseudocodex}

\usepackage{booktabs}
\usepackage{tablefootnote} 

\makeatletter
\g@addto@macro\bfseries{\boldmath}
\makeatother

\definecolor{labelkey}{rgb}{0,0.5,0.0}
\definecolor{royalpurple}{rgb}{0.47, 0.32, 0.66}

\usepackage{listings}
\definecolor{darkred}{rgb}{0.6,0.0,0}
\definecolor{darkgreen}{rgb}{0,0.4,0}

\usepackage{jheppub} 
\allowdisplaybreaks 

\usepackage[capitalise]{cleveref}

\definecolor{semiblue}{rgb}{0.3,0.3,0.8}
\newcommand{\logbook}[2]{}

\newcommand{\res}{\textsc{Powheg\,Box\,Res}}
\newcommand{\powheg}{\textsc{Powheg}}
\newcommand{\pythia}{\textsc{Pythia}}
\newcommand{\pythiaVer}{\textsc{Pythia8.3}}

\PassOptionsToPackage{hypertexnames=false}{jheppub}

\tikzstyle{block} = [rectangle, minimum width=1.0cm, minimum height=0.75cm, thin, draw=black]
\tikzstyle{blob} = [circle, minimum width=0.5cm, thin, draw=black]
\tikzset{blackarrow/.style={-stealth, semithick, draw=black}}
\tikzset{connection/.style={inner sep=0,outer sep=0}}

\renewcommand\tabularxcolumn[1]{m{#1}}
\newcolumntype{C}{>{\centering\arraybackslash}X}

\title{A Powheg generator for $t$- and $s$-channel single-top production and decay including non-resonant effects}

\preprint{\small Nikhef 2026-010, MS-TP-26-24}

\newcommand{\Torinoaff}{Dipartimento di Fisica, Università di Torino, and INFN, Sezione di Torino,
Via P. Giuria 1, I-10125 Torino, Italy}
\newcommand{\NIKHEFaff}{Nikhef, Theory Group, Science Park 105, 1098 XG, Amsterdam, The Netherlands}
\newcommand{\Munsteraff}{Institut für Theoretische Physik, Universität Münster, Wilhelm-Klemm-Stra\ss e 9, D-48149 Münster, Germany}

\author[a]{Melissa van Beekveld,}%
\author[a]{Robin van Bijleveld,}%
\author[b]{Silvia Ferrario Ravasio,}%
\author[c]{Tomáš Ježo,}%
\author[a]{Tommaso Saracco}%

\emailAdd{mbeekvel@nikhef.nl}
\emailAdd{rbijleve@nikhef.nl}
\emailAdd{silvia.ferrarioravasio@unito.it}
\emailAdd{tomas.jezo@uni-muenster.de}
\emailAdd{tsaracco@nikhef.nl}

\affiliation[a]{\NIKHEFaff}
\affiliation[b]{\Torinoaff}
\affiliation[c]{\Munsteraff}

\date{Received: date / Accepted: \today}

\abstract{
We present a next-to-leading order + parton shower generator for $t$- and $s$-channel single-top production in hadron colliders, for a leptonically-decaying top, including off-shell and non-resonant effects.
The generator, publicly available in the \res{} framework, describes the hadro-production of a $b$-jet and a leptonically decaying $W$-boson, accompanied by an additional jet, i.e.\ $pp\to \ell^{+} \nu_{\ell}\, j_b\, j $ and $pp\to \ell^{-} \bar{\nu}_{\ell}\, j_{\bar{b}}\, j $, where $j_b$ ($j_{\bar{b}}$) is a jet containing a bottom (anti)quark, and $j$ is a generic jet.
We also provide an interface to the \pythia{} parton shower, suited to matching in the presence of decaying resonances such as the top quark. 
Our matching is designed to combine the advantages of a five-flavour calculation with the description of bottom quark mass effects in final-state jets.
We perform a phenomenological study using fiducial cuts and discuss shower effects. 
}

\keywords{Top quark, LHC, NLO, Parton Shower
  \\[4em]
  \textit{For the purpose of Open Access, the authors have applied a CC BY
  public copyright licence to any Author Accepted Manuscript (AAM)
  version arising from this submission.}
}

\begin{document}

\maketitle

\section{Introduction}
\label{sec:introduction}
The top quark plays a central role in the physics programme of the Large Hadron Collider (LHC).
Precise measurements of top-quark production processes provide stringent tests of the Standard Model (SM), and at the same time can probe the existence of physics beyond the SM. 
While top quarks are produced predominantly in pairs through the strong interaction, the
electroweak interaction also allows the production of a single top quark through three standard production modes at leading order: the $t$-channel, mediated by the exchange of a virtual $W$ boson between a light-quark
line and a heavy-quark line; the $s$-channel, in which a time-like virtual $W$ boson decays to a
top quark and a bottom antiquark; and associated $tW$ production.
The single-top production process therefore provides direct access to the structure of the $Wtb$ vertex, the Cabibbo–Kobayashi–Maskawa (CKM) matrix element $V_{tb}$, top-quark polarisation effects, and possible anomalous couplings, making it an interesting process to measure.

\begin{figure}[t]
\centering
\begin{subfigure}[b]{.4\linewidth}
\includegraphics[width=1\textwidth,height=6cm,keepaspectratio,page=1]{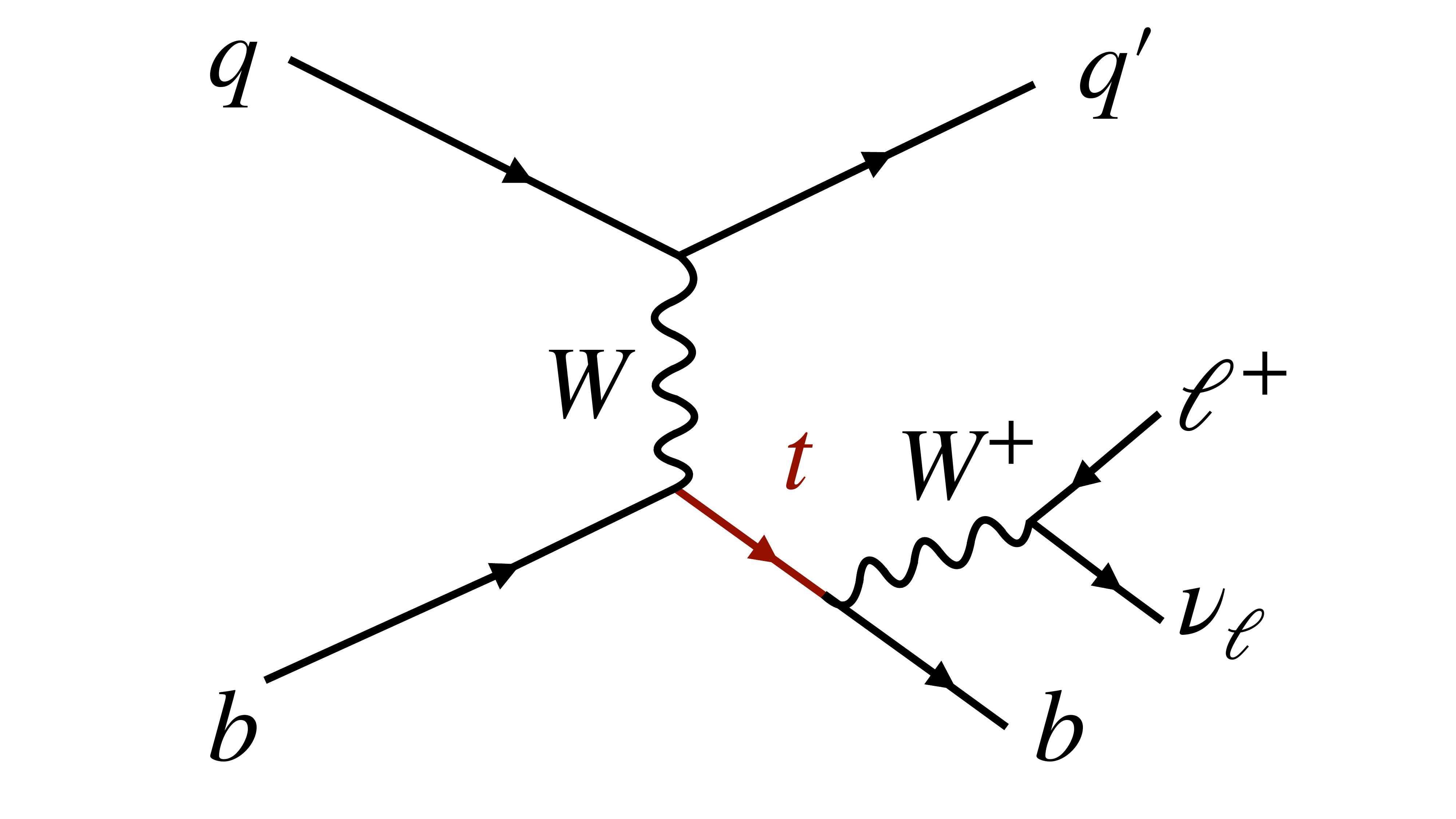}
\caption{}
\end{subfigure}
\hspace{1.5cm}
\begin{subfigure}[b]{.4\linewidth}
\includegraphics[width=1\textwidth,height=6cm,keepaspectratio,page=2]{Plots/diagrams.pdf}
\caption{}
\end{subfigure}
\caption{Examples of a $t$-channel (left) and $s$-channel (right) resonant diagram, with the top resonance highlighted in red.}
\label{fig:diagrams}
\end{figure}

At the LHC, the inclusive $Wbj$ final state is dominated by $t$-channel single-top production, for which $j$ is typically a forward light-flavour jet, while $s$-channel production contributes to the same final state when $j$ is a $\bar{b}$ jet. Examples of $t$- and $s$-channel resonant diagrams (with a leptonically-decaying $W$) are shown in Fig.~\ref{fig:diagrams}.
Single-top production was first observed at the Tevatron in 2009 by the D0 and CDF collaborations~\cite{D0:2009isq,CDF:2009itk}.
At the LHC, the ATLAS and CMS collaborations have measured inclusive and differential single-top cross sections over a range of different centre-of-mass (CM) energies~\cite{ATLAS:2023qrn,CMS:2012xhh,ATLAS:2017rso,CMS:2014mgj,CMS:2018lgn,CMS:2019jjp,ATLAS:2026vxj}.
Recent analyses reach a level of precision for which simulations incorporating higher-order QCD corrections, parton-shower effects, and off-shell contributions are mandatory.
From the theoretical side, single-top production has been studied extensively at next-to-leading order (NLO) in QCD, both in the narrow-width approximation but also including off-shell and non-resonant effects. 
An NNLO implementation of the $t$-channel single-top process in the narrow-width approximation is available in MCFM~\cite{Campbell:2020fhf}, and for the $s$-channel in Ref.~\cite{Liu:2018gxa}. 
At NLO, there are new partonic channels that belong to both production topologies. 
While a strict separation at the amplitude level is still possible, both production topologies contribute to the same physical final state and cannot be distinguished in a measurement.
Furthermore, beyond NLO, a separation between $t$- and $s$-channel at the amplitude level is no longer possible.\footnote{Moreover, the $s$-channel receives a contribution from $tW$ production in its real correction.}
Modelling single-top production and decay beyond the on-shell approximation, finite-width effects and the leading non-factorisable corrections connecting production and decay were first studied at NLO for the $t$-channel and subsequently for both the $t$- and $s$-channel processes within a systematic expansion about the top-quark pole~\cite{Falgari:2010sf,Falgari:2011qa}. 
For the $t$-channel, complete off-shell predictions, including resonant and non-resonant contributions and their interference, were later obtained at NLO in the five-flavour-number scheme (5FNS)~\cite{Papanastasiou:2013dta}, and extended to the SMEFT~\cite{Neumann:2019kvk}.

At the same time, modern collider analyses rely heavily on fully exclusive event generators matched to parton showers, which are capable of describing realistic hadronic final states.
Matching NLO calculations to parton showers is therefore essential for precision phenomenology.
Dedicated generators matched to parton showers (NLO+PS) for $s$- and $t$-channel single-top production were first constructed in the MC@NLO framework~\cite{Frixione:2005vw} and subsequently in \powheg~\cite{Alioli:2009je}. 
Both calculations employed on-shell production matrix elements in the 5FNS, with spin correlations in the subsequent top-quark decay included in the \powheg{} implementation. 
For $t$-channel production, a four-flavour-number scheme (4FNS) implementation, with massive bottoms, in both \powheg{} and aMC@NLO subsequently provided NLO accuracy also for observables involving the spectator $b$ jet~\cite{Frederix:2012dh}. 
Predictions for the $s$- and $t$- channels in both the 4FNS and 5FNS, supplemented by factorised spin-correlated top-quark decays, were later presented within SHERPA~\cite{Bothmann:2017jfv}, while a MiNLO-improved \powheg{} generator achieved simultaneous NLO accuracy for inclusive $t$-channel and $t$-channel-plus-jet observables~\cite{Carrazza:2018mix}.
Going beyond the on-shell approximation, the resonance-aware extension of the \powheg{} method was introduced and first demonstrated using a $t$-channel single-top process in \res{}~\cite{Jezo:2015aia}. 
A complete off-shell treatment of the $t$-channel production contribution in the 5FNS, including non-resonant contributions and their interference with the top-resonant contribution, was subsequently matched to the parton showers through the MC@NLO method~\cite{Frederix:2016rdc}. 
A later MadGraph5\_aMC@NLO study considered the complete off-shell prediction for a $t$-channel-enriched fiducial final state, including contributions from $s$-channel and $tW$ production, obtaining complete NLO QCD and electroweak predictions and separately studying the impact of NLO QCD parton-shower effects~\cite{Frederix:2019ubd}.

The \powheg{} method~\cite{Nason:2004rx, Frixione:2007vw, Alioli:2010xd} provides one of the standard frameworks for combining NLO QCD calculations with parton showers.
For processes involving intermediate resonances, a naive matching procedure can spoil the resonance virtuality and lead to distortions in observables sensitive to the resonance structure. 
This issue is particularly relevant for single-top production, where radiation can originate both from the production stage and from the decay of an intermediate top quark. 
The resonance-aware extension of the \powheg{} method implemented in the \res{} framework~\cite{Jezo:2015aia} addresses these issues by consistently associating radiation with specific resonance histories and by preserving resonance virtualities throughout the event generation. Since its advent it has been successfully applied to a range of processes~\cite{Jezo:2016ujg,Jezo:2023rht,Granata:2017iod,Chiesa:2019ulk,Buonocore:2022msy}.

In this work we present a new resonance-aware \res{}   generator for single-top production and decay, including both $t$- and $s$-channel contributions as well as non-resonant effects. 
The generator describes the leptonic final state $p\, p \to \ell^+\nu_\ell\, j_b\, j$
and its charge conjugate for single-antitop production. The Born contribution is of order $\mathcal{O}(\alpha^4)$ and NLO QCD corrections at $\mathcal{O}(\alpha_s\alpha^4)$. The generator includes all relevant resonant and non-resonant contributions and is matched to parton showers.
We work in the 5FNS, treating the $b$ quark as massless at the matrix-element level and including a non-zero $b$-quark PDF.
The jet label $j$ is defined inclusively with respect to flavour and can therefore also denote a $b$ jet.
The resulting final state contains both $t$-channel single-top production, for which $j$ is a light-flavour parton at Born level, and $s$-channel production, for which $j$ is a second $b$ quark (see Fig.~\ref{fig:diagrams}). 
Our set of real corrections also contains the LO contribution to $tW$-associated production in the 5FNS, as it yields the same final state. 
Our implementation consistently incorporates subprocesses that are formally non-resonant but required for infrared consistency in the \powheg{} framework. 
As we will show, some of these contributions are not only theoretically necessary, but can also become phenomenologically relevant after showering and fiducial event selection.
We provide a dedicated interface to the \textsc{Pythia} parton shower, implementing a resonance-aware veto strategy for the consistent treatment of emissions for every resonance structure.
The matching strategy is formulated for transverse-momentum-ordered showers and is designed to preserve their leading-logarithmic accuracy.

In the 5FNS employed here, the spectator $b$ quark in $t$-channel production first appears through real radiation and is therefore described only at LO accuracy, whereas a 4FNS calculation would include a massive spectator $b$ already at Born level and allow its kinematics to be predicted at NLO~\cite{Frederix:2012dh}.
Extending such a treatment to the complete off-shell process considered here, however, would entail a different calculation. 
The 5FNS, used here, is accurate to describe high-energy tails of observables that are not affected by bottom-quark mass effects. 
To smoothly transition to the regime where those effects do matter, we do introduce a novel hybrid treatment:
\powheg{} radiation from a bottom quark is restricted to transverse momenta above a scale of order $m_b$, while the softer region is delegated to \pythiaVer{}, where the finite $b$-quark mass regulates the collinear dynamics.
We pay particular attention to the mismatch between the massive and massless handover, making sure it happens above the scale of the bottom quark mass, as otherwise the leading-logarithmic accuracy of the prediction is lost. 

The paper is organised as follows. 
In Sec.~\ref{sec:code} we describe the implementation of the single-top process in the \res{} framework, including the treatment of resonance histories, the matching to the parton shower, and the handling of bottom-quark mass effects. 
In Sec.~\ref{sec:phenomenology} we present phenomenological results for fiducial single-top production at the LHC, focusing on the impact of non-resonant and $s$-channel contributions as well as on the modelling of QCD radiation by different shower algorithms. 
We summarise our findings and conclusions in Sec.~\ref{sec:conclusions}.

\section{\texorpdfstring{\powheg{} implementation}{Powheg implementation}}
\label{sec:code}
In this section, we provide details of the implementation of our code.
It is embedded in the \res~\cite{Jezo:2015aia} framework.
The matrix elements are generated with MG5\_aMC@NLO~\cite{Alwall:2014hca} in 5FNS and using the complex-mass scheme. 

The code is publicly available from the official \res{} \href{https://gitlab.com/POWHEG-BOX/RES}{GitLab repository} as a submodule linked to \url{https://gitlab.com/sferrario/single-top-dec-pwhg}. 
In the following, we discuss the process definition and the generation of the hardest emission, and provide details of the matching to \pythiaVer. Specific code flags are discussed in App.~\ref{sec:flags}.

\subsection{Process definition}\label{sec:process_definition}
We consider the hadronic process 
\begin{equation}
p\, p \to \ell^+\nu_\ell\, j_b\, j \, ,
\label{eq:processdef}
\end{equation}
together with the charge-conjugated process. Here, $j_b$ denotes a bottom-flavoured jet, i.e.\ it contains a $b$ quark, and $j$ denotes a generic QCD jet.
We work in 5FNS, so the generic jet may be initiated by a gluon or by any massless quark or anti-quark, including $b$ and $\bar b$.
The CKM matrix is taken to be block diagonal between the first two generations and the third generation, with $V_{tb}=1$, while mixing in the first two generations is retained.
The lowest perturbative order is $\mathcal{O}(\alpha^4)$.
At this order, the underlying Born flavour structures that describe the process in eq.~\eqref{eq:processdef} correspond to the first 3 entries of Tab.~\ref{tab:procs}, modulo permutations of the incoming partons.
Here and in the following, $u$ denotes either an up or charm quark, while $d$ denotes either a down or strange quark; CKM factors in the first two generations are understood.
We call a subprocess $t$-channel if its matrix element contains at least one electroweak boson exchanged in the $t$-channel, and $s$-channel otherwise. 
The generator can also simulate the charge-conjugate subprocesses of those reported in Tab.~\ref{tab:procs} for the corresponding single-antitop production and decay process in both the $t$- and $s$-channel.
\begin{table}[t] 
\centering 
\small
\begin{tabular}{c c c} 
\toprule
 & Subprocess & Channel\\
\midrule
$B_1$ & $u\, b \to \ell^+ \nu_\ell\, b\, d$ & $t$\\
$B_2$ & $\bar{d}\, b \to \ell^+ \nu_\ell\, b\, \bar{u}$ & $t$ \\
$B_3$ & $u\, \bar{d} \to \ell^+ \nu_\ell\, b\, \bar{b}$ & $s$ \\
$B_4$ & $u\, \bar{b} \to \ell^+ \nu_\ell\, \bar{b}\, d$ & $t$ \\
$B_5$ & $\bar{d}\, \bar{b} \to \ell^+ \nu_\ell\, \bar{b}\, \bar{u}$ & $t$ \\
$B_6$ & $b\, \bar{b} \to \ell^+ \nu_\ell\, \bar{u}\, d$ & $s$ \\
\bottomrule
\end{tabular}
\caption{Born flavour structures included in the calculation at $\mathcal{O}(\alpha^4)$ in the five-flavour scheme. The label in the last column denotes the $t$- or $s$-channel component used in the implementation.}
\label{tab:procs}
\end{table}

\begin{figure}[t]
    \centering
    \includegraphics[height=0.22\textheight,trim=149 35 260 44,clip,page=1]{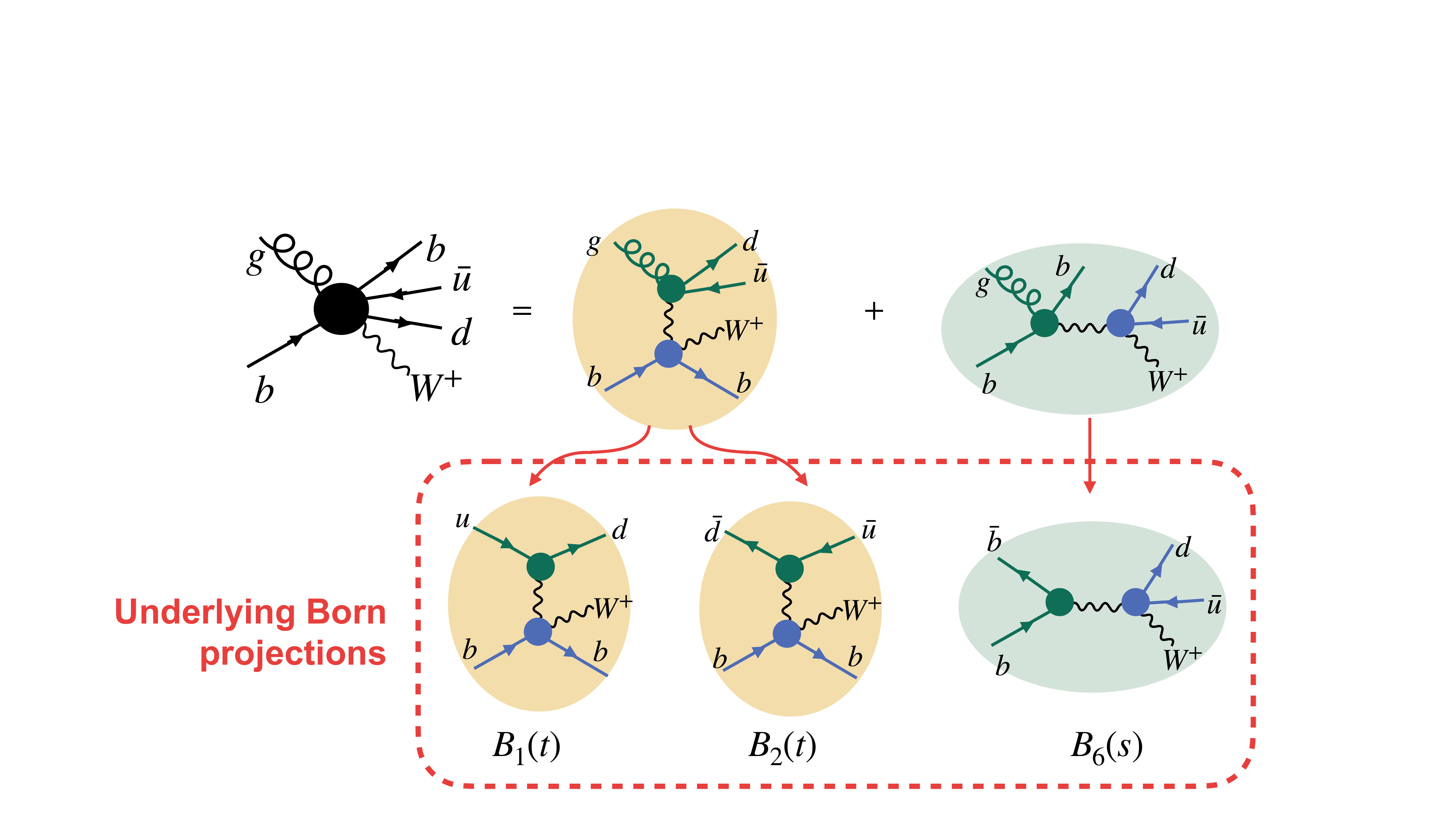}\hfill
    \includegraphics[height=0.22\textheight,trim=308 35 300 44,clip,page=2]{Plots/DiagramsUB.pdf}
    \caption{Representative real-emission flavour configurations entering the NLO corrections to single-top production and decay, shown with the final-state $W^+$ kept on shell for clarity: $g b \to W^+ d \bar u b$ (left) and $g u \to W^+ d \bar b b$ (right).
Such processes can be decomposed into $t$-channel (orange background) and $s$-channel (green background) contributions, whose interference vanishes by QCD colour algebra.
The lower diagrams show possible underlying Born configurations in the corresponding unresolved limits.
Different colours for the partons indicate different colour connections. The final-state $W^+$ boson can be attached either to a fermion line or to an intermediate gauge-boson propagator.
The full circles represent the underlying diagrammatic structures characterised by a common core topology; as such, the diagrams serve as a template for generating a full, gauge-invariant set of Feynman diagrams.}
    \label{fig:underlyingBorns}
\end{figure}

The subprocesses $B_1$ and $B_2$ are the usual resonant $t$-channel single-top Born configurations, while $B_3$ is the corresponding resonant $s$-channel configuration.
In these cases the top quark can go on shell and decay into a $b$ quark and an $\ell^+\nu_\ell$ pair.
The remaining Born structures, $B_4$, $B_5$ and $B_6$, are non-resonant.
In addition, $B_6$ is qualitatively different from the others, as it does not even correspond to the same final state at LO.
They are nevertheless required for the cancellation of infrared singularities in the NLO QCD corrections and are therefore included in our calculations.
This can be understood from Fig.~\ref{fig:underlyingBorns}, which shows two representative real-emission flavour configurations entering the NLO corrections to single-top production.
For example, the process $g b \to \ell^+\nu_\ell d\bar u b$ shown on the left is an incoherent sum of a $t$-channel contribution, which enters the real corrections to $B_1$ and $B_2$, and an $s$-channel contribution whose unresolved limit is associated with the non-resonant subprocess $B_6$, $b\bar b\to \ell^+\nu_\ell \bar u d$.
Similarly, the process $g u \to \ell^+\nu_\ell d b\bar b$, shown on the right, contains an $s$-channel contribution that factorises onto $B_3$ in the unresolved limit, and a $t$-channel contribution whose singular limits are associated with both $B_1$ and the non-resonant subprocess $B_4$.
The subprocess $B_5$ enters through the analogous light-quark flavour configurations with an incoming $\bar d$ instead of an incoming $u$.

At first sight one might expect that requiring a resolved final-state jet containing a $b$ quark, and not a $\bar b$ quark, removes the need for $B_4$, $B_5$ and $B_6$.
This can be true for a fixed-order calculation with such a flavour requirement, but in the \powheg{} method the event weight requires the real matrix element to be integrated over the full radiation phase space.
All singular regions that occur in this integral must therefore have a corresponding underlying Born contribution.
Moreover, current $b$-tagging techniques usually identify bottom-flavoured jets without distinguishing whether they contain a $b$ or a $\bar b$ quark.
The non-resonant subprocesses $B_4$ and $B_5$, which contain a final-state antibottom quark, can therefore pass a $t$-channel single-top selection.
The $b\bar b$-induced subprocess $B_6$ contains no final-state bottom quark at the hard Born level, but shower evolution can generate a final-state bottom-flavoured parton through backward evolution of an initial-state bottom quark via $g\to b\bar b$.
We find that, after showering, events originating from $B_6$ constitute the dominant Born--non-resonant background to $t$-channel single-top selection, making its inclusion phenomenologically important.\footnote{The role of $B_6$ is structurally analogous to that of photonless underlying Born configurations in FKS-based \powheg{} calculations of processes with an identified prompt photon~\cite{Barze:2014zba,Jezo:2016ypn,Jezo:2024wsc}.}
Such configurations are required to cover singular regions in which a photon is generated through collinear radiation, thereby providing the perturbative, pointlike component of parton-to-photon fragmentation.
Experimentally, its topology resembles electroweak $W+$jets production, whose real correction is dominated by a non-singular $tW$ production with a hadronic decay of the top. 
In experimental analyses the $tW$ production is considered a background and appropriately subtracted. 
However, not only is the inclusion of this contribution in the present generator required by a proper treatment of singularities, but incorporating it directly within the same event sample is actually required for a consistent propagation through showering and fiducial selection.

\subsection{Resonance-aware event generation}

We devote this section to the main features of our event generator embedded in the \res{} framework.
For efficient phase-space sampling, we generate the $t$- and $s$-channel components separately.
At the Born level this separation is straightforward, since each partonic subprocess in Tab.~\ref{tab:procs} belongs to either the $t$- or the $s$-channel component.
For real-radiation subprocesses containing diagrams of both types, such as those in Fig.~\ref{fig:underlyingBorns}, the interference between the $t$- and $s$-channel contributions vanishes by QCD colour algebra.
We can therefore perform the same separation at the matrix-element level.

For each chosen channel we keep all the resonance histories supported by the corresponding flavour structures. They are listed in Tab.~\ref{tab:resonances}.
\begin{table}[t] 
\centering 
\small
\begin{tabular}{l l} 
\toprule
$t$-channel & $s$-channel\\
\midrule
$t \to b \, (W^+\to \ell^+ \nu_\ell)$ 
& $W^+\to \bar{b}\, (t\to b (W^+\to\ell^+\nu_\ell))$ \\
$W^+\to \ell^+ \nu_\ell$ 
& $W^+\to (Z\to b\bar{b})\,(W^+\to \ell^+\nu_\ell) $\\
& $Z\to (W^+\to \ell^+\nu_\ell)\,(W^-\to q\,\bar{q}')$\\
\bottomrule
\end{tabular}
\caption{Implemented resonance structures for Born-level $t$- and $s$-channel.}
\label{tab:resonances}
\end{table}
In the $t$ channel, the possible histories either contain a leptonically-decaying top quark, as in $B_1$ and $B_2$, or contain no top resonance.
In the $s$ channel, the possible histories include a leptonically-decaying top quark, as in $B_3$, or a hadronically-decaying electroweak boson, as in the non-resonant configurations illustrated in Fig.~\ref{fig:underlyingBorns}.
For every subprocess, the \res{} framework uses kinematic projectors to select the most likely resonance history among the available ones, both for LO and NLO kinematics.
Here we rely on the original definition of the resonance history projectors rather than the more recent formulation of Ref.~\cite{Jezo:2023rht}.
For real-emission subprocesses that admit the same set of resonances as the underlying Born, the radiation is generated while preserving the chosen resonance structure.
In case a new resonance structure appears in the real-emission subprocess, the \res{} framework treats it as a separate regular region, i.e.\ as a finite tree-level contribution with an additional parton.
This is the case for the $s$-channel process $ g b \to \ell^+\nu_\ell \bar{u} d b$, which also admits a resonance history in which the final-state $\ell^+\nu_\ell b$ come from a leptonically-decaying top quark, which is not present for the Born process $b\bar b  \to \ell^+\nu_\ell \bar{u} d$.

For processes containing resonances, the \res{} framework can generate the hardest QCD emission in the production process and, independently, the hardest QCD emission in each coloured resonance decay.
Thus, $t$-channel events with a top-quark resonance can contain up to two emissions, one in production and one in the top decay, while events without a top resonance contain at most the production emission.
In the $s$ channel, an additional emission can be generated from a hadronically decaying electroweak boson whenever such a resonance history is present.
If both a top-quark resonance and a hadronically-decaying electroweak boson are present, the event can therefore contain up to three \powheg{} emissions.
This feature, made possible by the multiplicative nature of the \powheg{} method~\cite{Nason:2004rx}, ensures that the hardest emission off a resonance is described by the corresponding tree-level matrix element.
It also requires a dedicated interface to the subsequent parton shower, to avoid double counting emissions that are already present in the \powheg{} event.

\subsubsection{Matching with \pythia}
\label{sec:matching_py8}

\begin{figure}[t]
\setlength\abovecaptionskip{1pt}
\centering

\includegraphics[
  width=0.6\textwidth,
  page=1,
  trim={1.5cm 17.5cm 45cm 2.1cm},
  clip
]{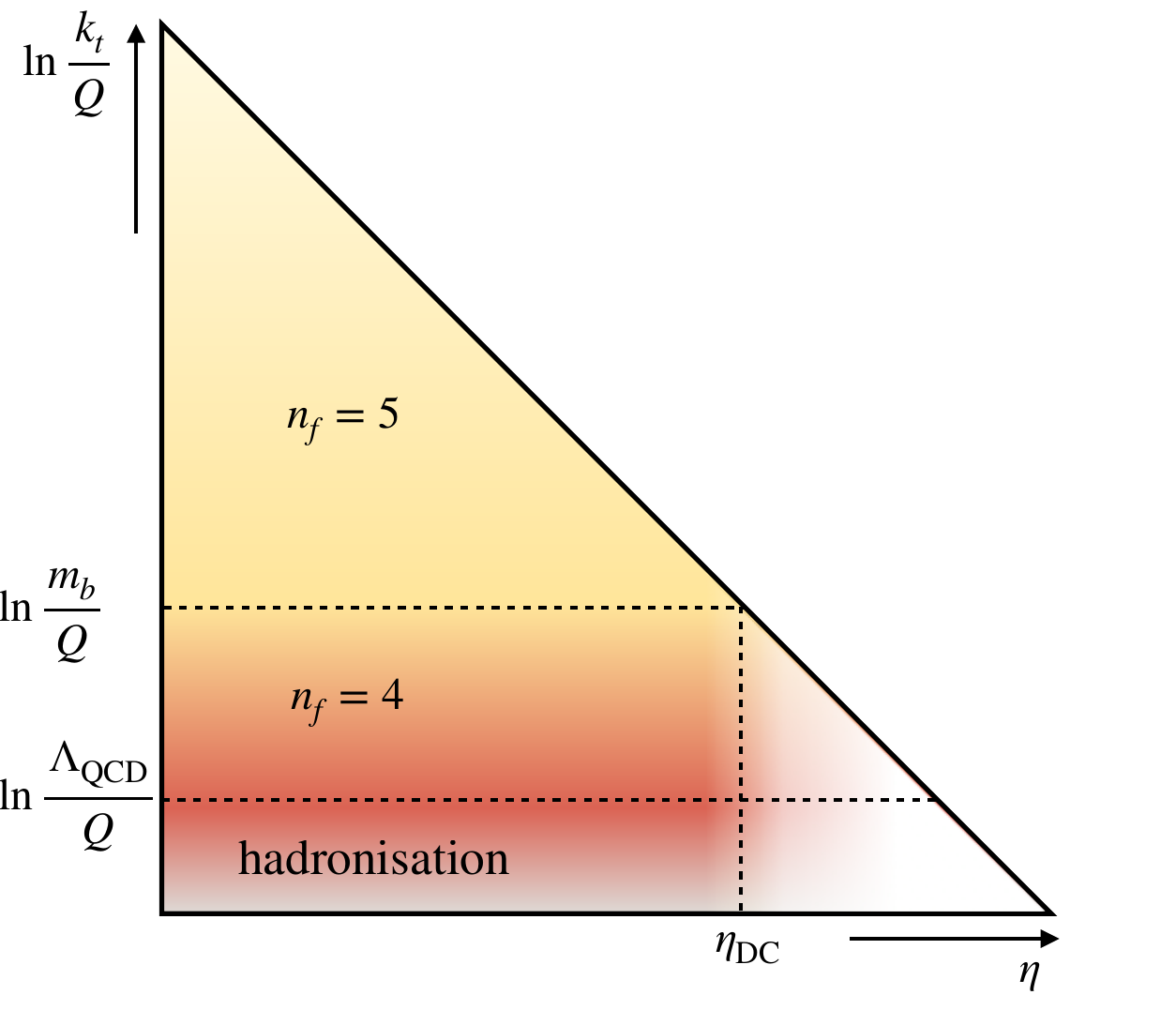}
\caption{Lund plane of the $b$-quark jet. We use $m_b$ to denote the bottom-quark mass, $Q$ for the hard scale and $\Lambda_{\rm QCD}$ for QCD cutoff scale.}
\label{fig:lundplane}
\end{figure}

We now describe how to match our event generator to \pythiaVer{}~\cite{Bierlich:2022pfr}, which is used for the subsequent parton shower.
The same veto strategy can be applied to any transverse-momentum ordered parton shower, preserving its leading-logarithmic accuracy.
Importantly, since we work in the 5FNS, the bottom quarks are considered massless in the calculation. 
However, when handing them over to  \pythiaVer{} for further showering, one must stop the generation of the hardest emission with \powheg{} \emph{above} the scale of the bottom quark mass. 
Not doing so will invalidate the leading-logarithmic accuracy of the shower~\cite{vanBeekveld:2026wnj}.
This can be seen easily in Fig.~\ref{fig:lundplane}, where we show the Lund plane~\cite{Andersson:1988gp} of the $b$-quark jet. 
For emissions with transverse momenta of the order of the mass of the $b$ jet, the emission starts feeling the effects of the deadcone at emission angles smaller than $\theta_{\rm DC}$ (or conversely, rapidities larger than $\eta_{\rm DC}$).
This effect is a leading-logarithmic effect.

According to the Les Houches event-file format~\cite{Alwall:2006yp}, every event carries a transverse-momentum scale, \texttt{SCALUP}.
In \powheg{} matching, \texttt{SCALUP} corresponds to the scale of the hardest emission generated by \powheg{}.
For processes without resonances, if the shower is ordered in transverse momentum, \texttt{SCALUP} can be used directly as the shower starting scale.
However, the effective transverse momentum used in the \res{} generator can differ significantly from the shower ordering variable for hard emissions.
We therefore start the shower from the highest allowed scale and, for each shower emission, compute its effective transverse momentum using the \powheg{} definition.
The emission is vetoed if this scale is larger than \texttt{SCALUP}.
In \pythiaVer{}, this veto is already implemented through the  \textsc{PowhegHooks} module.

For the present resonance-aware calculation we follow the strategy introduced in Ref.~\cite{FerrarioRavasio:2018whr}, where a new dedicated \textsc{PowhegHooks} was implemented to handle radiation from top-quark decay products in $pp\to t\bar{t}$.
In the presence of resonances, the parton shower first removes the resonance decays from the event record and showers the production process.
Then, starting from the innermost resonances, each resonance decay is restored and its decay products are showered.

Since \texttt{SCALUP} is set to the scale of the \powheg{} emission associated with the production process, the veto of subsequent production radiation is unchanged.
For events containing coloured resonance decays, we also compute a resonance-specific analogue of \texttt{SCALUP}, and use it as the veto scale when showering the corresponding decay.\footnote{The \res{} framework does not include resonance radiation in remnant events, i.e.\ events containing a very hard emission in production. For such events, the resonance-veto scale is set above the kinematic limit, leaving the shower unrestricted in resonance decays.}
If the event contains a decaying top quark, we denote this scale by $\zeta_t$. Its squared value is 
\begin{equation}
    \zeta^2_t = \begin{cases}
    \zeta^2_{0} & \mbox{ if }t\to Wb\\
  \displaystyle 
    \frac{2 (p_g \cdot p_t) (p_g \cdot p_b)}{(p_b \cdot p_t)} &\mbox{ if }t\to Wbg,
    \end{cases}
\end{equation}
where $\zeta_0$ is the infrared cutoff used by \res{} when generating radiation.\footnote{The value of $\zeta_0^2$ can be changed in the input file with the flag \texttt{ptsqmin}, see App.~\ref{sec:flags}, Tab.~\ref{tab:flagsres}.}
For a massless bottom quark, the second line corresponds to $\zeta^2_t = 2 E_g^2(1-\cos\theta_{bg})$, where $E_g$ and $\theta_{bg}$ are evaluated in the top rest frame.
For a $W$ boson decaying into a bottom-flavoured quark of momentum $p_b$ and a top or antitop quark, we compute
\begin{equation}
    \zeta^2_{W} = \begin{cases}
    \zeta^2_{0} & \mbox{ if }W \to t b\\
  \displaystyle 
    \frac{2 (p_g \cdot p_W) (p_g \cdot p_b)}{(p_b \cdot p_W)} &\mbox{ if }W\to tbg,
    \end{cases}
\end{equation}
where charge-conjugate decays are understood.
For a gauge boson $V$ (either $Z/\gamma^*$ or $W$) decaying into a light quark of momentum $p_q$ and a light antiquark of momentum $p_{\bar q'}$, we define
\begin{equation}
    \zeta^2_{V} = \begin{cases}
 \zeta^2_{0} &\mbox{ if }V \to q \bar{q}'\\
 \displaystyle
    2 (p_g \cdot p_V)\;\min\left(\frac{p_g \cdot p_q}{p_q \cdot p_V} , \frac{p_g \cdot p_{\bar q'}}{p_{\bar q'}\cdot p_V} 
    \right)&\mbox{ if }V \to q \bar{q}' g,
    \label{eq:Vhadr}
    \end{cases}
\end{equation}
i.e.\ when radiation is present, we compute the transverse momentum with respect to both light-quark directions in the resonance rest frame, and use the smallest value.
For a resonance $r$ that decays leptonically, we set $\zeta^2_r$ above the kinematic limit, so that no shower emission is vetoed in that decay.

The resonance-veto scale is labelled and looked up by the signed particle-id ancestry of the resonance, obtained by following its mother chain up to the hard-process root and retaining the status-22 resonance entries. This ancestry-based labelling is independent of \pythiaVer’s event ordering and shower-system index, and allows the scale to be matched consistently to the corresponding resonance decay.

In summary, an initial-state shower emission is rejected if its effective \powheg{}-like transverse momentum is larger than \texttt{SCALUP}.
The same veto is applied to final-state emissions that do not originate from a resonance decay.
For final-state radiation from the decay products of a resonance $r$, either an electroweak boson or a top quark, the associated resonance scale $\zeta^2_r$ serves as the veto scale instead.
We apply the veto only to QCD emissions, since our generator includes QCD corrections only.

\subsubsection{Recovering bottom-quark mass effects}
\label{sec:mass_effects}
For processes with bottoms in the initial state, the five-flavour scheme is the most suited one, because it allows to resum the logarithms of the bottom mass that appear in inclusive quantities such as the total cross section.
This choice also allows to resum large Sudakov logarithms in a region where the transverse momentum of the $b$ jet is small compared to the centre-of-mass energy, but quite above the bottom mass.
However, bottom-quark mass effects can be relevant when modelling the top-quark decay.
Therefore, one might wish to combine the massless and massive description of bottom quarks to take advantages of both.

As customary in NLO or LO  event generators, although the hard matrix elements are evaluated in the five-flavour scheme with massless bottom quarks, the Les Houches events~\cite{Alwall:2006yp} contain massive bottom and charm quarks, as well as massive leptons.
In particular, the event kinematics is reshuffled after the \powheg{} hardest-emission generation in order to assign these masses while preserving the resonance structure.
Subsequent emissions generated by the parton shower, which implements massive splitting kernels, therefore include the corresponding bottom-mass effects.

For radiation with transverse momentum above the bottom mass $m_b$, mass effects are small, and the matrix elements implemented in our generator yield a more accurate description than the splitting kernel implemented in the shower.  
On the other hand, for radiation with transverse momentum of order $m_b$ or below, it is preferable to let the shower generate the emission with its massive splitting kernels, rather than to use the exact tree-level matrix element in the massless approximation.
In \pythiaVer{}, these kernels are also supplemented by matrix-element corrections for example for radiation in top decays.
We therefore set the \res{} cutoff for the hardest emission generation to $\zeta_0=m_b=4.8$~GeV.
This means that the hardest emission from the production process and from each resonance decay are generated by \res{} using the exact massless tree-level matrix element when harder than $m_b$, otherwise are generated by the parton shower with massive splitting functions.

Let us inspect at the level of the cross section what this means.
For simplicity, we consider a single radiating system and retain at most the hardest emission.
We write the massless \powheg{} weight as
\begin{equation}
  \bar B_0(\Phi_n) =
  B_0(\Phi_n) + V_0(\Phi_n)
  + \int d\Phi_{\rm rad}\, R_0(\Phi_{n+1}),
\end{equation}
where $B_0$, $V_0$ and $R_0$ are the Born, virtual and real contributions evaluated with massless bottom quarks.
The corresponding cross section is
\begin{align}
  d\sigma_{\textsc{pwg}}
  = \bar B_0(\Phi_n) d\Phi_n
  \Bigg[
    \Delta_0(k_{t}^{\max},m_b)
    + \frac{R_0(\Phi_{n+1})}{B_0(\Phi_n)}
      \,\Theta(k_t-m_b)
      \,\Delta_0(k_{t}^{\max}, k_t) \,d\Phi_{\rm rad}
  \Bigg],
  \label{eq:pwgxsec}
\end{align}
where $k_t=k_t(\Phi_{\rm rad})$ is the transverse momentum of the emission and
\begin{equation}
  \Delta_0(k_{t}^{\max}, k_t) =
  \exp\left[
    -\int d\Phi_{\rm rad}'\,
    \frac{R_0(\Phi_{n+1}')}{B_0(\Phi_n)}
   \, \Theta(k_t'-k_t)\, \Theta(k_t^{\rm max}-k_t')
  \right]
\end{equation}
is the \powheg{} Sudakov.
Expanding the Sudakov to first order in $\alpha_s$, the \powheg{} cross section can be written as the sum of a Born-like contribution and a contribution with one resolved emission:
\begin{subequations}
  \begin{align}
    \frac{d\sigma_{\textsc{pwg}}^{(n)}}{d\Phi_n}
    &= \bar B_0(\Phi_n)\Delta_0(k_{t}^{\max},m_b) \nonumber \\
    &= B_0(\Phi_n) + V_0(\Phi_n)
      + \int d\Phi_{\rm rad}\,
        R_0(\Phi_{n+1})\, \Theta(m_b-k_t)
      + \mathcal{O}(\alpha_s^2), \\
    \frac{d\sigma_{\textsc{pwg}}^{(n+1)}}{d\Phi_{n+1}}
    &= \bar B_0(\Phi_n)
      \frac{R_0(\Phi_{n+1})}{B_0(\Phi_n)}
      \, \Theta(k_t-m_b)\Delta_0(k_{t}^{\max}, k_t) \nonumber \\
    &= R_0(\Phi_{n+1})\, \Theta(k_t-m_b)
      + \mathcal{O}(\alpha_s^2).
  \end{align}
\end{subequations}
Thus emissions above $m_b$ are described by the exact massless real matrix element, while the unresolved part of the massless real correction remains in the Born-like contribution.

We now consider the action of the parton shower on a Born configuration.
When no \powheg{} emission is generated, the relevant shower starting scale is $m_b$.
Keeping at most one shower emission gives
\begin{align}
  d\sigma_{\textsc{ps}}
  = B_0(\Phi_n)d\Phi_n
  \Big[
    \Delta_m(m_b,q_0)
    + P_m(\Phi_{\rm rad})\, \Theta(m_b-k_t)\, \Theta(k_t-q_0)
      \Delta_m(m_b,k_t)d\Phi_{\rm rad}
  \Big],
\end{align}
where $q_0$ is the shower infrared cutoff, $P_m(\Phi_{\rm rad})$ is the massive shower emission kernel and $\Delta_m$ is the corresponding shower Sudakov.
Defining the shower approximation to the real cross section as
\begin{equation}
  R_m^{\textsc{ps}}(\Phi_{n+1})
  = B_0(\Phi_n)P_m(\Phi_{\rm rad}),
\end{equation}
its expansion reads
\begin{align}
  d\sigma_{\textsc{ps}}
  =& \left[
      B_0(\Phi_n)
      - \int d\Phi_{\rm rad}\,
        R_m^{\textsc{ps}}(\Phi_{n+1})
        \, \Theta(m_b-k_t)\, \Theta(k_t-q_0)
    \right]d\Phi_n \nonumber \\
  &+ R_m^{\textsc{ps}}(\Phi_{n+1})
    \, \Theta(m_b-k_t)\, \Theta(k_t-q_0)d\Phi_{n+1}
    + \mathcal{O}(\alpha_s^2).
\end{align} 
Combining this expansion with the \powheg{} result and taking the formal limit $q_0\to 0$, one obtains
\begin{align}
  d\sigma_{\textsc{nlops}}
  =& \left[
      B_0(\Phi_n) + V_0(\Phi_n)
      + \int d\Phi_{\rm rad}\,
        \left(R_0(\Phi_{n+1})-R_m^{\textsc{ps}}(\Phi_{n+1})\right)
        \, \Theta(m_b-k_t)
    \right]d\Phi_n \nonumber \\
  &+ \left[
      R_m^{\textsc{ps}}(\Phi_{n+1})\, \Theta(m_b-k_t)
      + R_0(\Phi_{n+1})\, \Theta(k_t-m_b)
    \right]d\Phi_{n+1}
    + \mathcal{O}(\alpha_s^2).
\end{align}
The integral in the first line corresponds to an effective massification of the virtual amplitude, in which one removes the collinear divergences by adding the integrated massless term $R_0$, and replaces them with mass logarithms originated from integrating $R_m$.
For inclusive observables, one can integrate over the radiation phase space and the dependence on $R_m$ drops, recovering the standard massless result, while exclusive observables can retain the dominant bottom-mass effects generated by the shower.
This prescription preserves the advantages of the five-flavour scheme, such as the resummation of initial-state collinear logarithms in the total cross section, while recovering approximate bottom-mass effects in more exclusive observables.
A similar strategy has been used in \textsc{PanScales} to combine showers with massive partons and massless QCD NLO calculations~\cite{vanBeekveld:2026wnj}.

\begin{figure}[t!]
    \centering
    \includegraphics[width=0.55\linewidth]{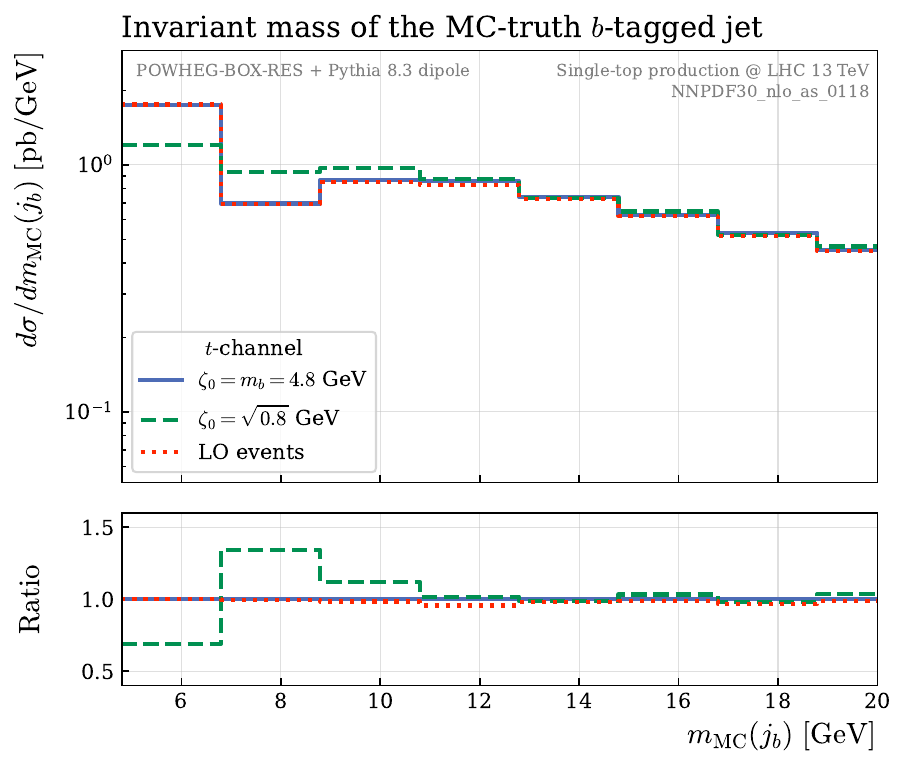}
    \caption{Comparison for different values of the \res{} cutoff $\zeta_0$, for $m_b=4.8$~GeV (blue, solid), the default value $\sqrt{0.8}$~GeV (green, dashed), and LO events (red, dotted).}
    \label{fig:mb}
\end{figure}

To assess the quality of our hybrid five-/four-flavour description of top decay, we proceed as follows. We generate LO events, reshuffle their momenta so that they contain massive heavy quarks, and then retain only events containing a top resonance, and shower them with \pythiaVer{}.
We use \pythiaVer{} to approximate the massive calculation, as it includes finite-mass effects and matrix-element corrections for radiation in top decay.
We then repeat the same procedure, but allow \res{} to generate the hardest emission before massifying the momenta and passing the event to the shower.
Events produced in this manner have still LO normalisation.
The resulting distributions are shown in Fig.~\ref{fig:mb}.
The observable displayed is the invariant mass of the system comprising all partonic top-decay products and therefore corresponds to a Monte-Carlo-truth definition of the $b$-jet mass.
This idealised observable allows us to disentangle QCD corrections associated with the production mechanism, for which the 5FNS description is more accurate, and to focus specifically on the modelling of the top decay, where finite-mass effects are important.
For the generation of the hardest emission, we consider both the case in which the bottom-quark mass is used as the cutoff for the \res{} generator and the case in which the default value, $\zeta_0=\sqrt{0.8}$~GeV, is used.
We find that using $m_b$ as the cutoff in \res{} yields a distribution compatible with that obtained using the massive tree-level matrix-element corrections implemented in \pythiaVer{}. This demonstrates that the treatment of real radiation in top decay correctly incorporates finite-$m_b$ effects.
By contrast, the default cutoff $\zeta_0=\sqrt{0.8}$~GeV does not provide a correct description of the dead-cone region, as is visible for $m_{\rm MC}(j_b)<10$~GeV. 
On the other hand, the hard-tail modelling is very similar for all setups.

We note that one could, in principle, raise the \res{} cutoff above $m_b$ and still obtain a similar description of radiation in top decay. However, this would degrade the modelling of hard radiation associated with the production mechanism, where \pythiaVer{} does not provide matrix element corrections.
Furthermore, we have verified that, for inclusive distributions in which bottom-mass effects are small, cutoff values in \res{} ranging from approximately $1$~GeV to $\mathcal{O}(m_b)$ give compatible results. This is expected because, at small transverse momentum, the parton shower kernel is a good approximation of the exact matrix element.
For progressively larger cutoff values, however, the uncertainty of hard radiation in the production process increases.
Therefore in the following we adopt the bottom mass as cutoff for generating radiation in \res{}.
One might notice that when computing the equivalent \powheg{}-like hardness for an emission, as defined in Sec.~\ref{sec:matching_py8}, the expressions assume massless quarks.
Large corrections may arise for very small transverse momenta, but for scales that are typically much smaller than $m_b$, leaving the outcome of the veto procedure mostly unaffected.

All distributions shown in the main body of the paper are calculated using the resonance scales as defined in Sec.~\ref{sec:matching_py8}, and with a cutoff of $\zeta_0 = m_b$. 
In App.~\ref{sec:hooks} we study an alternative method to veto emissions.
The difference between the prescription of Sec.~\ref{sec:matching_py8} and that of App.~\ref{sec:hooks} may be considered as an additional theoretical uncertainty.


\section{\texorpdfstring{Phenomenology for $t$-channel}{Phenomenology for t-channel}}
\label{sec:phenomenology}
For the phenomenological analysis, we consider a centre-of-mass energy of 13 TeV and we used the NNPDF30\_nlo\_as\_0118\ NLO PDF set \cite{NNPDF:2014otw, Buckley:2014ana}. We set the top mass to $m_t =  173.2$~GeV and the NLO top width to be $1.3569$~GeV \cite{Papanastasiou:2013dta}. For simplicity, we switch off the damping functionality (both \texttt{hdamp} and \texttt{bornzerodamp}), but the code supports it.\footnote{With damping enabled (\texttt{hdamp 173.2} and \texttt{bornzerodamp 1}) the remnant cross section is only about 2\% and 1\% in $t$- and $s$-channel samples, respectively.}
We select a minimal setup of cuts, namely the ones that are common to Refs.~\cite{ATLAS:2026vxj,ATLAS:2024ojr,CMS:2020vac,CMS:2019jjp}. We require at least two jets with $p_T>30$~GeV. We cluster the jets via the anti-$k_t$ algorithm \cite{Cacciari:2008gp}, as implemented in \textsc{FastJet} \cite{Cacciari:2011ma} with $R=0.4$. We demand to have exactly one $b$-tagged jet, with the tagging being performed with the interleaved flavour-neutralisation (IFN) algorithm \cite{Caola:2023wpj}. We require the $b$ jet to have pseudorapidity $|\eta(j_b)|<2.5$.
For the leptonic final state we demand that $|p_T(\ell)| > 28~\rm{GeV}$ and a missing transverse energy of $E_{T}^{\rm{miss}} > 30~\rm{GeV}$. 
Furthermore we require the $W$ boson to have a transverse mass $m_T(W) > 50~\rm{GeV}$, which is defined as $m^2_T(W) = 2p_T(\ell)E_T^{\rm{miss}}(1-\cos\phi(\vec{p}_T^{\,\,\rm{miss}},\ell))$. 
Lepton isolation cuts are not implemented in this analysis.
We employ the default simple shower, but with the \texttt{SpaceShower:dipoleRecoil} option~\cite{Cabouat:2017rzi}. In Sec.~\ref{sec:shower} we illustrate why this option is necessary for modelling additional QCD radiation.
We run without a QED shower, and we turn off multi-parton interactions and hadronisation effects. 
This set of cuts constitutes a minimal definition for the `$t$-channel' fiducial phase space.

An estimate of the scale uncertainty is performed by means of the 7-point scale variation, i.e.\ varying the renormalisation and factorisation scales around a central scale $\mu_0$, in the range
\begin{equation}
	\label{eq:7-pt-var}
	\biggl( \frac{\mu_R}{\mu_0}, \frac{\mu_F}{\mu_0} \biggr) \in \bigl\{ (0.5,0.5), (0.5,1), (1,0.5), (1,1), (1,2), (2,1), (2,2) \bigr\}.
\end{equation}
In the following, the central scale is chosen to be $\mu_0=H_T/2$, with $H_T= \sum_{i=1}^n\sqrt{m_i^2 + p_{T,i}^2}$ and $i$ running over the final-state particles.  The \res{} code allows the scale choice to depend on the resonance history, but for convenience, we choose it same for all resonance histories.

In Sec.~\ref{sec:background} we will discuss the role of $s$-channel single-top production as opposed to $t$-channel. 
We also remind the reader that our $s$-channel contribution contains the $b\bar{b}$-initiated subprocesses, which also incorporates the associated $tW$ production at LO, which yields the same final state as real corrections to the $t$-channel single-top production and therefore is not distinguishable. 
We will show its importance for certain observables, and show some relevant NLOPS-accurate observables. In Sec.~\ref{sec:shower}, we illustrate why the default recoil option in \pythia{} leads to the wrong radiation pattern, and one should instead enable the fully-local option (\texttt{SpaceShower:dipoleRecoil}).

\subsection{Channel composition of the fiducial region}\label{sec:background}
In this section we aim to assess the impact of $s$-channel  contributions compared to those in the $t$-channel. In the case of studying $t$-channel single top in an experimental setup, rather than relying on removing the $s$-channel via fiducial cuts, it is more accurate to simulate both $t$- and $s$-channel separately.

The fiducial selection described above enriches the $t$-channel contribution but also accepts events from the $s$-channel component. 
We quantify their impact here by comparing the $t$-channel sample with the sum of the seperately generated $t$- and $s$-channel samples. 
Here, the $s$-channel component is defined as in Sec.~\ref{sec:process_definition} and includes the $B_6$ sector, whose real-emission contributions contain associated $tW$ production at LO.

Tab.~\ref{tab:xsec} shows that including this component raises the fiducial top-production cross section from $3.25$~pb to $4.25$~pb, i.e.\ an increase of approximately $30\%$.
For antitop production, the corresponding increase is larger at approximately $46\%$.
The $s$-channel component therefore remains substantial after fiducial selection. 
After implementing more stringent cuts on the fiducial phase space as done in the experimental setup of Ref.~\cite{ATLAS:2026vxj} (not shown in Tab.~\ref{tab:xsec}), the $s$-channel contribution remains $14\%$. 
\begin{table}[t] 
\centering 
\small
\begin{tabular}{l c c } 
\toprule
 & $t$-ch. [pb] & $t$-ch.+$s$-ch. [pb]  \\
\midrule
single top  & $3.25^{+0.11}_{-0.07} $ &  $4.25^{+0.19}_{-0.14} $ 
\vspace{1ex}
\\
single antitop & $1.98^{+0.07}_{-0.04} $ & $2.89^{+0.15}_{-0.11} $ \\
\bottomrule
\end{tabular}
\caption{Fiducial cross section for $t$-channel and the combined $t$- and $s$-channel, both for single-top and single-antitop production. Uncertainty is computed with seven-point scale variation. Cuts are defined at the beginning of Sec~\ref{sec:phenomenology}.}
\label{tab:xsec}
\end{table}
This showcases that cuts will not remove this component, and that it is important to model it in any experimental analyses that targets $t$-channel single-top quark production. 

\begin{figure}[t]
\centering
\begin{subfigure}[b]{.45\linewidth}
\includegraphics[width=1\textwidth,height=6cm,keepaspectratio]{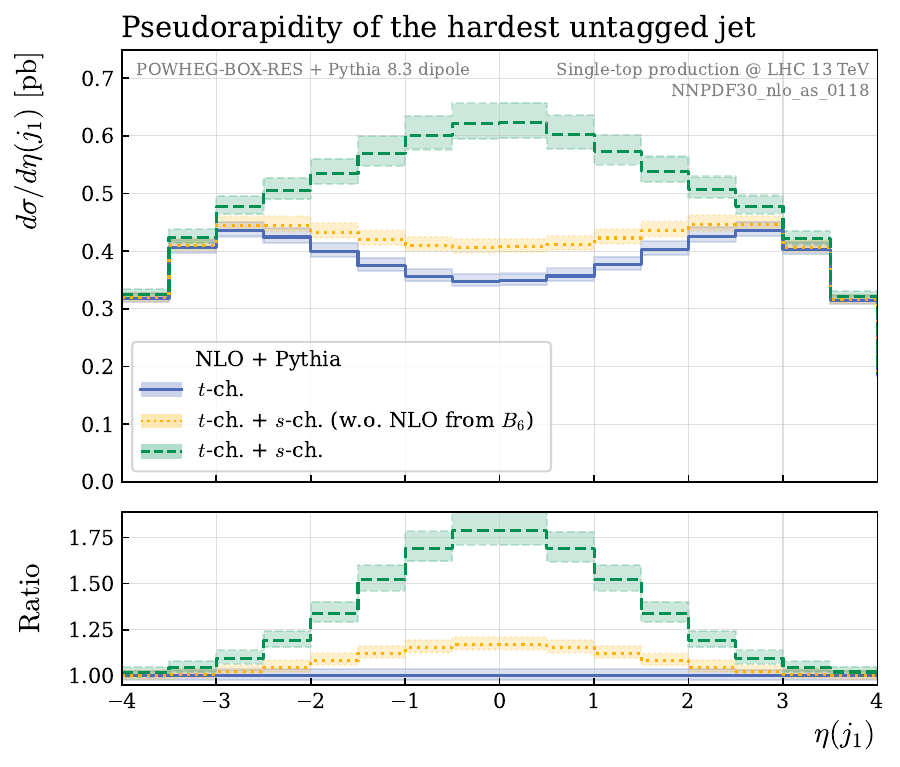}
\caption{}
\end{subfigure}
\hspace{0.5cm}
\begin{subfigure}[b]{.45\linewidth}
\includegraphics[width=1\textwidth,height=6cm,keepaspectratio]{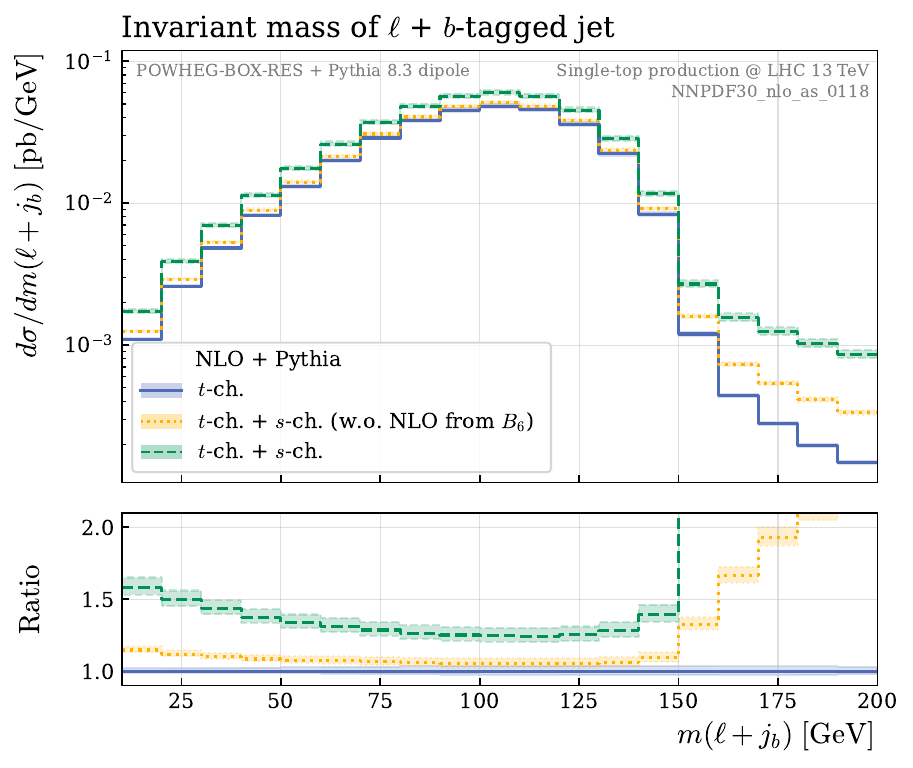}
\caption{}
\end{subfigure}
\caption{Plots for the pseudorapidity of the hardest untagged jet (left) or invariant mass of the $\ell$+$b$-jet system (right), where we compare $t$-channel (blue, solid) and combined $t$- and $s$-channel contributions with (green, dashed) and without (yellow, dotted) $b\bar{b}$ contributions, i.e.\ NLO processes coming from the $B_6$ Born subprocess showed in Tab.~\ref{tab:procs}, which are dominated by $tW$ production. Error bands indicate the seven-point scale variation uncertainty.}
\label{fig:t-vs-ts}
\end{figure}

The impact of the additional component on differential distributions is illustrated in Fig.~\ref{fig:t-vs-ts} for top production. 
The pseudorapidity $\eta$ of the hardest untagged jet, $j_1$, is enhanced most strongly in the central region.
There, the $t$-channel contribution has a natural suppression (also see Sec.~\ref{sec:shower}), whereas the $s$-channel contribution is unsuppressed. 
The real emission process that have $B_6$ as an underlying Born process (see Tab.~\ref{tab:procs}), are dominated by $tW$ production. 
These topologies are normally not included when considering `$t$-channel' fiducial phase space for single-top production, but we want to stress again that they are needed for the consistency of the calculation. 
This process does not have the rapidity suppression of the hardest untagged jet in the central rapidity region, and it fills up the central rapidity region. 
Applying the more stringent cuts of Ref.~\cite{ATLAS:2026vxj} (not shown), will still lead to an enhancement in the central region of about $40\%$, which again primarily comes from these $tW$-resonant contributions. 
The invariant mass of the lepton-plus-$b$-jet system, $m_{\ell + j_b}$, is also affected, particularly in the high-mass tail. 
The sharp drop near $m_{\ell + j_b} \sim \sqrt{m_t^2 - m_W^2}$ is created by the on-shell top-decay endpoint. 
Off-shell, non-resonant and radiative effects populate the region above it. 
This explains why the $s$-channel contribution, which relatively has more diagrams without a top resonance than the $t$-channel contribution, populates that region more strongly. 
Other observables, not shown here, that are affected in shape by the additional $s$-channel contributions, are for example the rapidities and transverse momenta of the lepton, the $b$-jet and untagged subleading jets.  

\begin{figure}[t]
\centering
\begin{subfigure}[b]{.45\linewidth}
\includegraphics[width=1\textwidth,height=6cm,keepaspectratio]{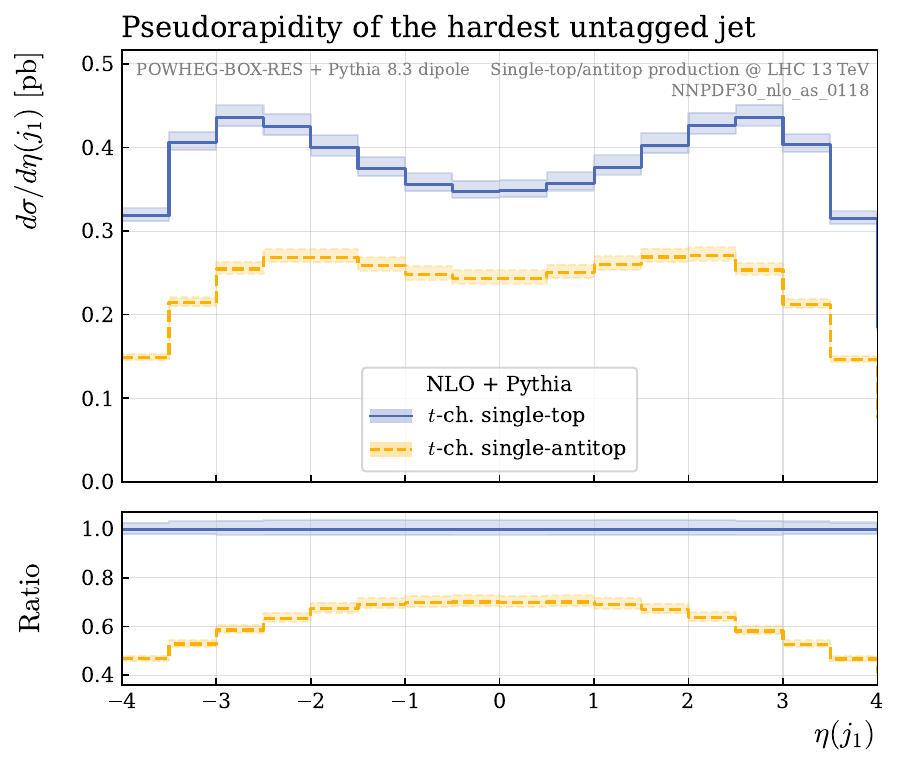}
\caption{}
\end{subfigure}
\hspace{0.5cm}
\begin{subfigure}[b]{.45\linewidth}
\includegraphics[width=1\textwidth,height=6cm,keepaspectratio]{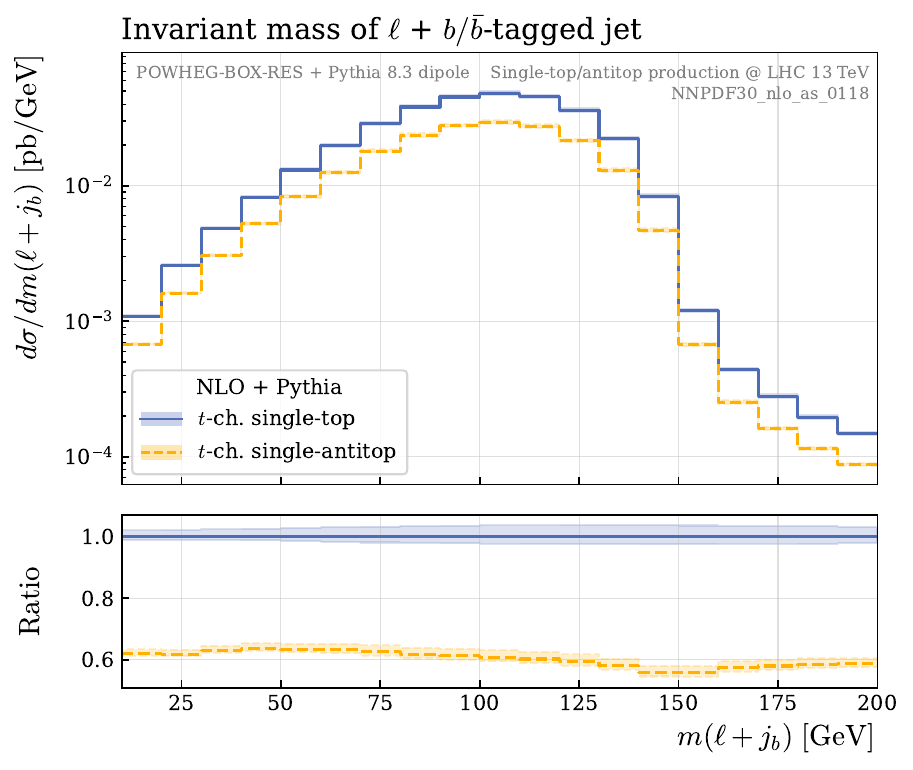}
\caption{}
\end{subfigure}
\caption{
Plots for the pseudorapidity of the hardest untagged jet (left) or invariant mass of the charged lepton/antilepton and the $b$/$\bar{b}$-tagged jet system (right) for single-top (blue, solid) or single-antitop production (yellow, dashed). Error bands indicate the seven-point scale variation uncertainty.}
\label{fig:top-vs-antitop}
\end{figure}

We finish this subsection by inspecting the difference between top and antitop for purely the $t$-channel contributions, as shown in Fig.~\ref{fig:top-vs-antitop}. 
The larger top-production rate reflects the dominant contribution from valence-$u$-initiated subprocesses, whereas antitop production is more sensitive to the smaller $d$-quark distribution. 
The difference becomes more pronounced at large $|\eta_{j_1}|$, where larger partonic momentum fractions are probed. 
Consequently, the antitop-to-top ratio shown in the lower panel decreases towards large $|\eta_{j_1}|$. 
Conversely, the $m_{\ell + j_b}$ distribution is only affected by an overall normalisation.


\subsection{Shower effects}\label{sec:shower}

\begin{figure}[t]
    \centering
    \includegraphics[width=1.\linewidth]{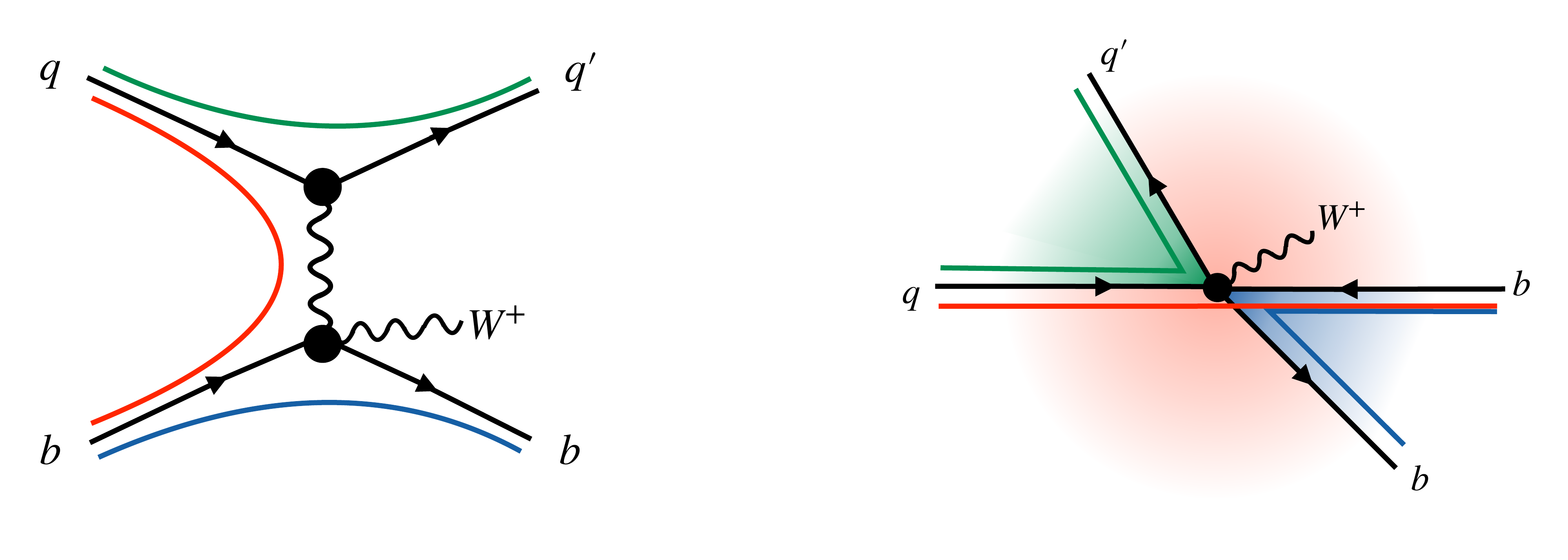}
    \caption{Typical diagram for $t$-channel production of a top quark at LO, which subsequently decays into a $b$-quark and a $W^+$-boson (left on shell in the picture for clarity), shown both as a Feynman diagram (left) and as it appears in the centre-of-mass frame (right). The two colour dipoles present at LO are shown in green and blue, while the spurious unphysical \((qb)\) dipole modelled by the simple \pythia{} shower with default recoil scheme is shown in red.}
    \label{fig:LOtchan}
\end{figure}

In this section, we address another important feature of processes such as \(t\)-channel single-top production, vector-boson fusion, and vector-boson scattering: they can be viewed as two independent deep-inelastic scattering (DIS) processes, each involving a pair of colour-connected partons~\cite{Ballestrero:2018anz, Jager:2020hkz, Hoche:2021mkv}. 
Within a dipole-shower framework, this implies that, at LO, radiation can originate only from the two dipoles spanned by colour-connected partons. For a typical LO topology, such as that shown in Fig.~\ref{fig:LOtchan}, these are the dipoles formed by the \((qq')\) and \((bb)\) pairs.
Colour coherence strongly suppresses emissions outside the region spanned by each dipole. Consequently, if the two final-state jets are back-to-back, widely separated in rapidity, and close to the beam axes, we expect a gap in the radiation pattern in the central-rapidity region.
Thus, in the following results we impose, in addition to the fiducial cuts described at the beginning of Sec.~\ref{sec:phenomenology}, the additional requirements
\begin{equation}
    |\eta(j_b)-\eta(j_1)|>2.5 \;,\quad \eta(j_b)\cdot\eta(j_1)<0 \;,
\end{equation}
where $j_b$ is the $b$-tagged jet and $j_1$ is the hardest untagged jet.

\begin{figure}[t]
    \centering
     \includegraphics[width=0.45\linewidth]{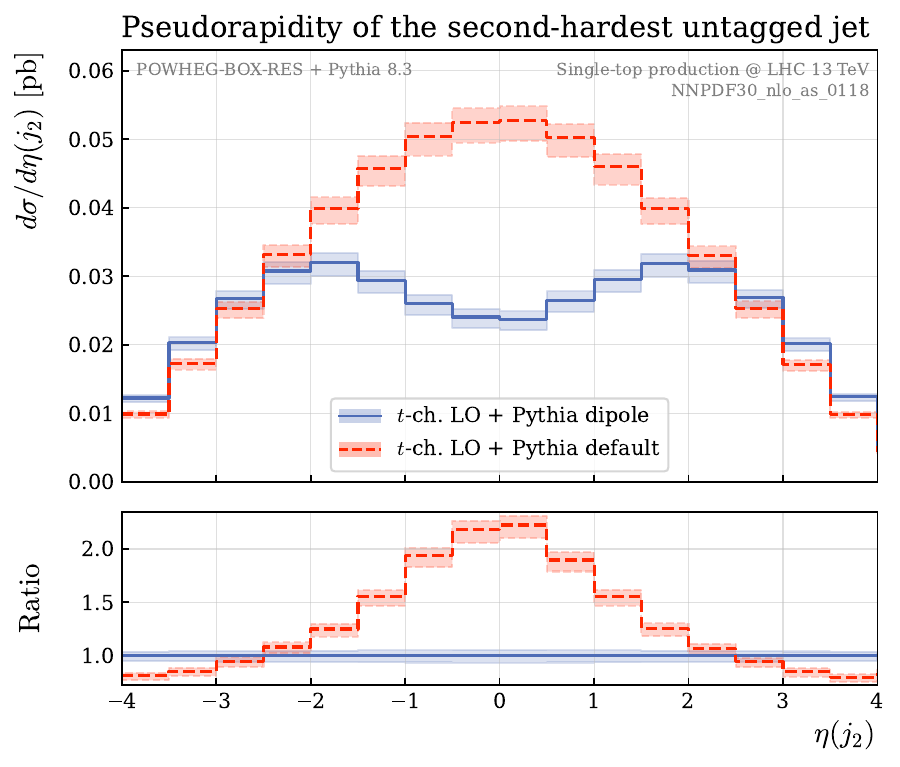}
    \caption{Distribution of the second hardest light jet for
     LO events showered with the simple \pythia{} shower, with the default global recoil~(red) as well as the dipole-local one~(blue). Error bands indicate the seven-point scale variation uncertainty. 
    }
    \label{fig:etalj2LO}
\end{figure}

We begin by showering LO events with \pythia{}'s simple shower, using either the default (global) recoil option for initial-state radiation,
as well as the dipole-local one, which is activated by the flag \texttt{SpaceShower:dipoleRecoil}. The latter corresponds to the choice we have adopted in the previous sections.
The problem of \pythia{}'s default recoil strategy is that, for initial-state emissions, it employs a fictitious initial-initial dipole, which in the example of Fig.~\ref{fig:LOtchan} would be formed by the $(qb)$-pair, and generates soft emissions within the whole angular distribution, violating colour coherence. 
On the contrary, \pythia{}'s fully local map uses the momentum of the final-state parton in the emitting dipole to limit the angular region of initial-state radiation.
The impact of the different treatment of initial-state radiation is clearly visible by looking at the kinematic distributions of first jet generated by the shower (i.e.~the second-hardest jet that is not $b$-tagged) in Fig.~\ref{fig:etalj2LO}. 
Here, we show the pseudorapidity distribution of the second-hardest light jet, i.e.~the first shower-generated jet, where we shower the LO events with \pythia{}'s simple shower using the default global recoil, as well as using the dipole one.
As expected, the dipole-local option suppresses radiation in the central region, while the default one fills it almost completely.\footnote{For illustrative purposes we tuned down the LO starting scale \texttt{SCALUP} of the shower. At LO, it is by default set to the centre-of-mass energy, whereas the events are produced with a starting scale of the $p_T$ of the fixed-order final-state light quark. A starting scale at the centre-of-mass energy would induce emissions that are much higher than the Born kinematics would allow.}  

\begin{figure}[t]
    \centering
    \begin{subfigure}[]{.45\linewidth}
    \includegraphics[width=\linewidth]{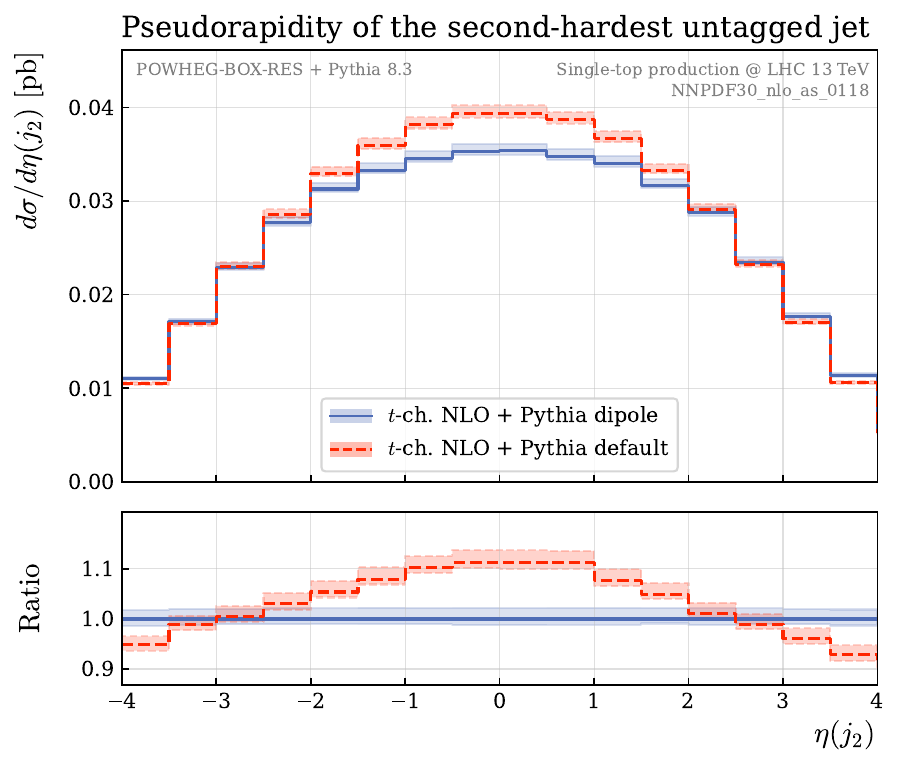}
    \caption{}
    \label{fig:etalj2NLO}
    \end{subfigure}
    \hfill
    \begin{subfigure}[]{.45\linewidth}
    \includegraphics[width=\linewidth]{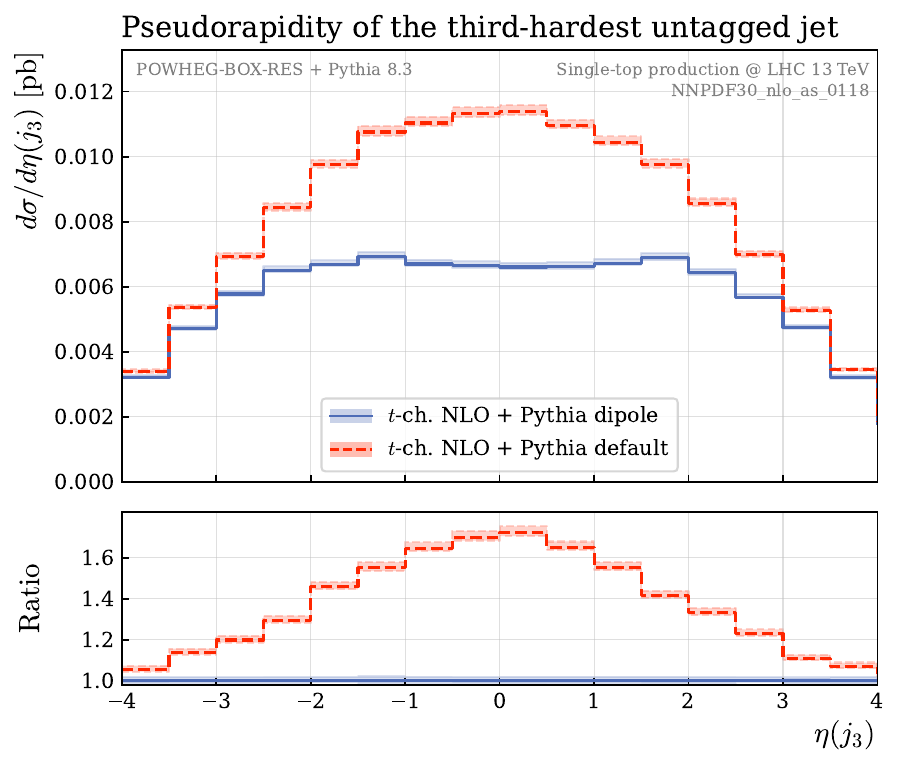}
    \caption{}
    \label{fig:etalj3NLO}
    \end{subfigure} 
    \caption{Distribution of the second- (left) and third-hardest (right) light jet for 
  NLO events showered with the simple \pythia{} shower, with the default global recoil~(red) as well as the dipole-local one~(blue). Error bands indicate the seven-point scale variation uncertainty.}
\end{figure}

We now move to showering NLO events. 
Now, the first emission is modelled according to the exact matrix element that is implemented. 
Because of this, the shower will have minimal impact on the kinematics of the second harder jet, meaning that the $\eta$ distribution of this jet will be similar between the default and the fully-local dipole option in \pythia{}. 
This is indeed the case as can be seen in Fig.~\ref{fig:etalj2NLO}. Note that a residual difference of roughly 10\% remains between the two showers, consistent with the behaviour already seen for VBF in Ref.~\cite{Hoche:2021mkv}. 
Nevertheless, the problem is not solved, but just moved to the next hardest jet.
The discrepancy due to the failure of \pythia{}'s default recoil option to preserve colour coherence is now shifted to the next-hardest jet: now the third-hardest untagged jet will be the first one produced by the shower. Fig.~\ref{fig:etalj3NLO} confirms that the same issue persists there.
It is therefore imperative to use the dipole recoil option when using \pythia's simple shower in order to retain the correct modelling of soft radiation.

\section{Conclusions}
\label{sec:conclusions}
In this work, we have presented a new resonance-aware \textsc{Powheg-Box-Res} generator for the processes
$pp \to \ell^+\nu_\ell j_b j$ and its charge conjugate. 
The calculation combines NLO QCD accuracy with parton showers and includes $t$- and $s$-channel single-top production,
top decay, and the associated off-shell and non-resonant contributions. 
Associated $tW$ production at LO naturally arises in the set of real corrections.
Radiation in production and in each resonance decay is generated independently while preserving the resonance virtualities.
We have also developed an interface to \pythiaVer, which consistently vetoes shower emissions in the production and decay processes. 
The resulting public generator can be downloaded from the official \res{} \href{https://gitlab.com/POWHEG-BOX/RES}{GitLab repository}, as a submodule linked to \url{https://gitlab.com/sferrario/single-top-dec-pwhg}. 

The calculations are performed in the five-flavour scheme, where bottom quarks are massless in the hard matrix elements and are proton constituents. 
To recover the finite-mass effects, we introduced a novel hybrid prescription in which \res{} generates emissions only above the scale of the bottom-quark mass, while softer radiation is delegated to the shower,
which treats final-state bottom quarks as massive. 
This retains the correct LL resummation of both the initial-state bottom quark logarithms while incorporating the correct mass effects.

Our phenomenological study demonstrates that contributions beyond the resonant $t$-channel signal remain important after realistic fiducial cuts. 
Adding the $s$-channel contribution increases the fiducial cross section by roughly $30\%$, and a contribution of about $14\%$ remains under tighter cuts. 
The effects can be even larger in differential distributions, reaching up to $40\%$ in the central region of the pseudorapidity of the light jet. 
Formally non-resonant subprocesses required by IR consistency also survive event selection. 
This shows that fiducial cuts do not isolate the $t$-channel contribution and that the full final state should be simulated consistently. 

Finally, we have examined the sensitivity of additional jet radiation to the shower recoil prescription. 
It is well known that the default \pythia{} shower introduces unphysical colour connections that lead to an overpopulation of the central-rapidity region, unless the dipole-local recoil strategy~\cite{Cabouat:2017rzi} is employed. 
The choice of the correct recoil strategy is particularly relevant especially for the $t$-channel topology.

\acknowledgments
We thank Rikkert Frederix and Valentin Hirschi for providing the relevant files in order to reproduce their results from Ref.~\cite{Papanastasiou:2013dta}. 
We also thank Paolo Torrielli and Emanuele Re for useful discussions.
This research was supported
by the Italian Ministry of Universities and Research (MUR) under the
FIS grant (CUP:D53C24005480001, FLAME) (SFR),
by the Dutch Research Council (NWO) under project number
VI.Veni.232.190 (MvB).
We thank each other's institutions for the hospitality during different stages of this work.

\newpage

\appendix

\section{Input card options}
\label{sec:flags}
In Tab.~\ref{tab:flagsproc} we list the process-specific flags for the single-top generator, together with their default values and a brief description of their function.
In Tab.~\ref{tab:flagsres} we report \textsc{Powheg} flags that are particularly relevant for the single-top process, and, if different from the default, recommended values for them.
In particular, the flag \texttt{allrad} has to be set to \texttt{1} to take maximum advantage of the resonance-aware features of the code, \texttt{iupperfsr} has also to be set to \texttt{1} to be able to handle radiation from $W\to q \bar{q}$ in the event generation.
The flag \texttt{for\_reweighting} can be set to \texttt{1} to include the virtual corrections only via reweighting during the event generation, making the event generation considerably faster.
The parameters \texttt{ptsqmin}, \texttt{charmthr} and \texttt{bottomthr} control the thresholds for the radiation in the event generation from light quarks, charm and bottom quarks, respectively. We recommend the charm and the bottom thresholds to be set to the square of the charm and bottom masses in the previous table, or to slightly higher values, to allow for the inclusion of mass effects via parton showering.

Furthermore, we list the flags that are relevant for showering the events with \pythia{}, using our \texttt{\texttt{Powheg}HooksST} class, which is derived from the \texttt{\texttt{Powheg}Hooks} class in \pythia{}.  We report their recommended values in Tab.~\ref{tab:flagspythia}.

\begin{table}[th!]
\centering
\small
\renewcommand\tabularxcolumn[1]{p{#1}}
\begin{tabularx}{\linewidth}{l l l >{\raggedright\arraybackslash}X}
\toprule
Flag & Type & Default & Description \\
\midrule
\texttt{process} & \texttt{<int>} & \texttt{1} & $t$-channel (\texttt{1}) or $s$-channel (\texttt{2}) single-top production. \\
\texttt{conjugate} & \texttt{<int>} & \texttt{0} & Top (\texttt{0}) or antitop (\texttt{1}) production. \\
\texttt{scalechoice} & \texttt{<int>} & \texttt{0} & Central value for the renormalisation and factorisation scales \\
& & & \texttt{0}: $\mu_0=m_t/2$; \\
& & & \texttt{1}: $\mu_0=H_T/2$; \\
& & & \texttt{2}: $\mu_0=Q/2$; \\
& & & \texttt{3}: $\mu_0=H_T(\text{partons})+p_T(\ell^+)$.\\
\texttt{alphaem\_inv} & \texttt{<double>} & \texttt{132.50698} & $1/\alpha_{\rm EW}$. \\
\texttt{gfermi} & \texttt{<double>} & \texttt{1.16639d-5} & $G_F$ [GeV$^{-2}$]. \\
\texttt{tmass} & \texttt{<double>} & \texttt{172.5} & $m_t$ [GeV].\\
\texttt{zmass} & \texttt{<double>} & \texttt{91.188} & $m_Z$ [GeV]. \\
\texttt{twidth} & \texttt{<double>} & \texttt{1.5083} & $\Gamma_t$ [GeV]. \\
\texttt{zwidth} & \texttt{<double>} & \texttt{2.441} & $\Gamma_Z$ [GeV]. \\
\texttt{wwidth} & \texttt{<double>} & \texttt{2.0476} & $\Gamma_W$ [GeV] \\
\texttt{bmass} & \texttt{<double>} & \texttt{4.80} & $m_b$ [GeV] in events. \\
\texttt{cmass} & \texttt{<double>} & \texttt{1.5} & $m_c$ [GeV] in events. \\
\texttt{verbose} & \texttt{<int>} & \texttt{0} & If \texttt{1}, print MG5 NLO stability checks. \\
\texttt{diagonalCKM} & \texttt{<int>} & \texttt{0} & If \texttt{1}, use diagonal CKM matrix. \\
\texttt{CKM\_Vij} & \texttt{<double>} & Ref.~\cite{ParticleDataGroup:2024cfk} & Control individual CKM matrix elements, with \texttt{ij} equal to \texttt{ud}, \texttt{us}, \texttt{cd}, or \texttt{cs}. \\
\texttt{collinearcut} & \texttt{<double>} & \texttt{0.01} & Avoid singularity in virtual matrix elements by setting them to zero if a final-state quark has $p_T$ lower than \texttt{collinearcut}. \\
\texttt{emscales} & \texttt{<int>} & \texttt{0} & See App.~\ref{sec:hooks}. \\
\bottomrule
\end{tabularx}
\caption{Process-specific flags for the single-top generator.}
\label{tab:flagsproc}
\end{table}

\begin{table}[t]
\centering
\small
\renewcommand\tabularxcolumn[1]{p{#1}}
\begin{tabularx}{\linewidth}{l l l l>{\raggedright\arraybackslash}X}
\toprule
Flag & Type & Default & Recommended& Description \\
\midrule
\texttt{allrad} & \texttt{<int>} & \texttt{0} &\texttt{1} & If \texttt{1}, include all possible radiations in the event generation, otherwise only the hardest one. \\
\texttt{iupperfsr} & \texttt{<int>} & \texttt{0}& \texttt{1} & Upper bound for FSR radiation, (1) is the only one that supports a massless recoiler.\\
\texttt{for\_reweighting} & \texttt{<int>} & \texttt{0}& \texttt{1} & If \texttt{1}, include the virtual corrections only via reweighting during the event generation.\\
\texttt{ptsqmin} & \texttt{<double>} & \texttt{0.8} & $m_b^2$ & Minimum $p_T^2$ [GeV$^2$] for radiation in the event generation. \\
\bottomrule
\end{tabularx}
\caption{Default and recommended \textsc{Powheg} flags that are particularly relevant for the single-top process.}
\label{tab:flagsres}
\end{table}

\begin{table}[t]
\centering
\small
\renewcommand\tabularxcolumn[1]{p{#1}}
\begin{tabularx}{\linewidth}{l l l >{\raggedright\arraybackslash}X}
\toprule
\texttt{Powheg:<flag>}
 & Type & Recommended & Description \\
\midrule
\texttt{veto} & \texttt{<int>} & \texttt{1} & If \texttt{1}, veto emissions in the parton shower above the scale of the \texttt{Powheg} radiation, both in production and resonances decay. \\
\texttt{nfinal} & \texttt{<int>} & \texttt{0} & If \texttt{0}, do not rely on the number of final-state partons after stripping resonances.\\
\texttt{vetoCount} & \texttt{<int>} & $\geq \texttt{5}$ & Perform the veto until a certain number of emissions are accepted. Count separately production and resonance decay emissions.\\
\texttt{pThard} & \texttt{<int>} & \texttt{0} & If \texttt{0}, use \texttt{SCALUP} as veto scale for radiation in production.\\
\texttt{pTemt} & \texttt{<int>} & \texttt{0} & If \texttt{0}, check the transverse momentum of the emitter w.r.t the radiator (for both production and resonance decay).\\
\texttt{emitted} & \texttt{<int>} & \texttt{0} & If \texttt{0}, use \texttt{Pythia} definition of emitter and emitted partons (for both production and resonance decay).\\
\texttt{pTdef} & \texttt{<int>} & \texttt{0} & If \texttt{0}, use \texttt{\texttt{Powheg}} definition of transverse momentum in production (this is always used in resonance decay).\\
\texttt{QEDveto} & \texttt{<int>} & \texttt{0} & If \texttt{0}, do not veto QED emissions above the \texttt{Powheg} scale.\\
\texttt{pTcutoff} & \texttt{<double>} & $\sqrt{\texttt{ptsqmin}}$ &
Used instead of \texttt{SCALUP} for resonance decays into two particles, when no additional radiation is present. [GeV]. \\
\bottomrule
\end{tabularx}
\caption{Recommended \texttt{\texttt{Powheg}HooksST} flags for showering the events with \pythia{}.}
\label{tab:flagspythia}
\end{table}

\section{Discussion on veto scale in hooks}
\label{sec:hooks}
Here we describe an extension of the \res{} framework developed to validate our \pythiaVer{} \texttt{UserHook} and to assess systematically the effect of reconstructing resonance-decay veto scales from reshuffled massive $b$-quark momenta rather than from the original massless \powheg{} kinematics.

In the standard workflow, \res{} communicates the shower starting scale for production radiation through the \texttt{SCALUP} variable of the LHE interface. 
No analogous standard entry exists for the scales associated with radiation from resonance decays. 
These scales, which are required to ensure a consistent handover from \powheg{} to the parton shower, must therefore be reconstructed event by event by the shower interface.

In \pythia, this reconstruction is commonly implemented through the \texttt{UserHook} machinery. 
For example, the hook~\cite{FerrarioRavasio:2018whr} distributed with the \texttt{bb4l} generator~\cite{Jezo:2016ujg} provides two alternative strategies based on \texttt{doVetoFSREmission} and \texttt{scaleResonanceVeto}, respectively. 
The former inspects each trial final-state emission and vetoes it whenever its hardness, evaluated according to the \powheg{} definition, exceeds that of the corresponding \powheg{} emission. 
In \texttt{bb4l}, the relevant emitter can be identified unambiguously and the resonance rest frame can be reconstructed from the event record. 
Consequently, the \powheg{} emission scale can be recovered accurately from the LHE kinematics.

The situation is more involved for the process considered here. 
\powheg{} generates the radiation using five-flavour-scheme kinematics, with massless $b$ quarks, whereas the momenta are subsequently reshuffled to assign a finite mass to the $b$ quarks before the event is written out to the LHE file. 
Reconstructing the veto scale from these modified momenta therefore does not reproduce exactly the scale obtained from the original radiation kinematics. 
Moreover, the identity of the \powheg{} emitter is not communicated by the standard LHE record, which can introduce an additional ambiguity in hadronic resonance decays. 
This happens for example in the hadronic decay of a gauge boson into two light quarks, in which case our default \texttt{UserHook} strategy selects the smallest possible scale, see Eq.~\eqref{eq:Vhadr}.

Both presciptions are NLO+LL accurate, and the two will generate differences beyond this accuracy. 
Their numerical impact is a priori unknown.
In the region where they can have a large numerical impact it is not even clear which prescription is more suitable. 
That is, is it better if the shower starting scale is calculated in the massless kinematics in \powheg{} because this is how the emission was attached, or better if this scale is calculated using massive kinematics in \pythia{} because the subsequent showering will rely on massive kinematics.
To assess the relevance of this mismatch, we extended \res{} to retain the veto scales computed directly from the native \powheg{} radiation kinematics. 
For every generated production or resonance-decay emission, the modified \powheg{} executable stores the emission hardness together with the flavours of the emitter and radiated parton and the index of the associated resonance. 
This information is written into the corresponding LHE event as comment lines and can subsequently be read by the shower interface. 
The extension is optional and is activated through the \powheg{} input flag \texttt{emscales}.

\subsection{The \texttt{emscales} block}

For each emission, the stored information consists of the \powheg{} hardness
\begin{equation}
  \zeta_r = \sqrt{t}\,,
  \label{eq:emscale}
\end{equation}
where $t$ is the \powheg{} ordering variable, i.e.~the radiated transverse momentum squared in
the generation of the hardest emission~\cite{Alioli:2010xd} in the rest frame of the resonance $r$. 
The record further contains the PDG identifiers of the emitter and radiated parton, \texttt{id\_emitter} and \texttt{id\_radiated}, respectively. 
The variable \texttt{ires} identifies the resonance system with which the emission is associated. 
Production-stage emissions are assigned \texttt{ires=0}, while radiation generated in a resonance decay carries the index of the corresponding resonance in the event record.
Thus, \texttt{id\_emitter} and \texttt{id\_radiated} specify the flavour structure of the branching, whereas \texttt{ires} specifies the production or resonance-decay system to which its veto scale belongs.
Schematically, the additional record reads
\begin{verbatim}
#emscales nemissions=N_em
#emscales i=1 scalup=pT_em(1) id_emitter=... id_radiated=... ires=0
#emscales i=2 scalup=pT_em(2) id_emitter=... id_radiated=... ires=nres(2)
...
#emscales i=N_em scalup=pT_em(N_em) id_emitter=... id_radiated=... ires=nres(N_em)
\end{verbatim}
In the \texttt{allrad} setup, the record can contain at most one production emission and at most one emission associated with each resonance, such that $N_{\mathrm{em}}\leq N_{\mathrm{res}}+1$. The stored list is reset at the beginning of each event. The corresponding \powheg{} reader routines also recognise the \texttt{emscales} block, allowing the information to be preserved when LHE events are read and modified (e.g.~for reweighting).

\subsection{\texttt{UserHook} using \texttt{emscales}}

Our \pythia{} interface uses a single \texttt{PowhegHooksST.h} implementation for both the standard and the \texttt{emscales}-based matching prescriptions. 
The latter is activated by setting \texttt{emscales 1} in the \texttt{powheg.input} file.%
\footnote{At present, \texttt{emscales 1} cannot be set through the \pythia{}'s \texttt{.cmnd} file.} 
In this mode, the interface parses the \texttt{emscales} lines and associates each stored \powheg{} emission scale with the corresponding production or resonance-decay shower system. 
As for the standard case, the resonance-veto scale is then labelled by the signed particle-id ancestry of the resonance.
The veto then proceeds exactly as in the standard \texttt{doVetoFSREmission} implementation, except that the reference scale is taken directly from the stored \powheg{} information rather than reconstructed from the Les Houches event. 

\begin{figure}[t]
\centering
\begin{subfigure}[b]{.45\linewidth}
\includegraphics[width=1\textwidth,height=6cm,keepaspectratio]{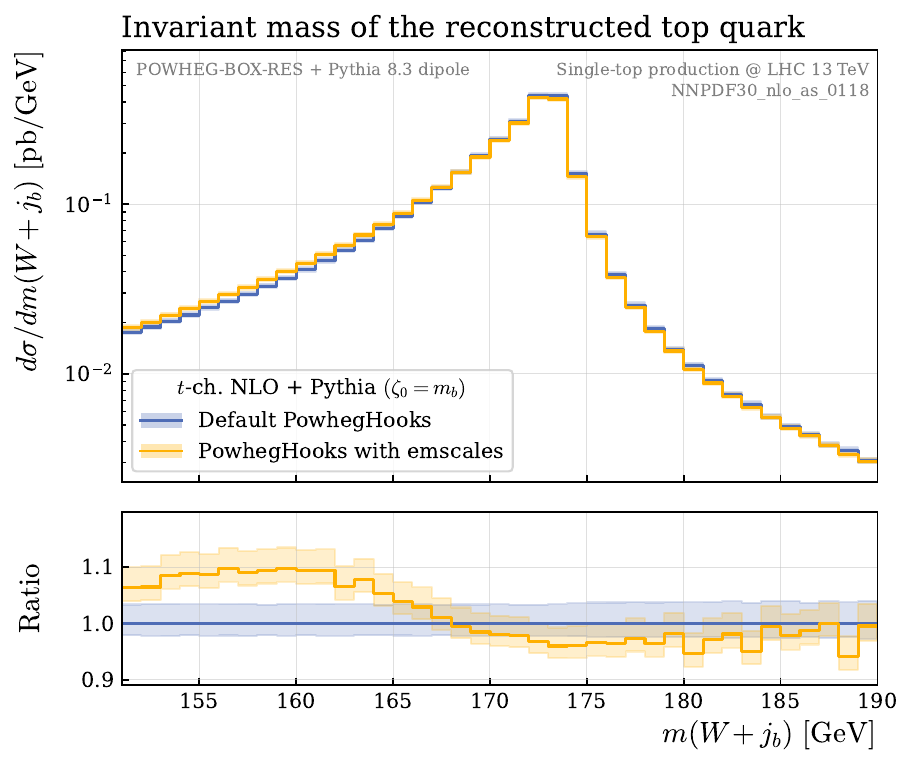}
\caption{}
\end{subfigure}
\hfill
\begin{subfigure}[b]{.45\linewidth}
\includegraphics[width=1\textwidth,height=6cm,keepaspectratio]{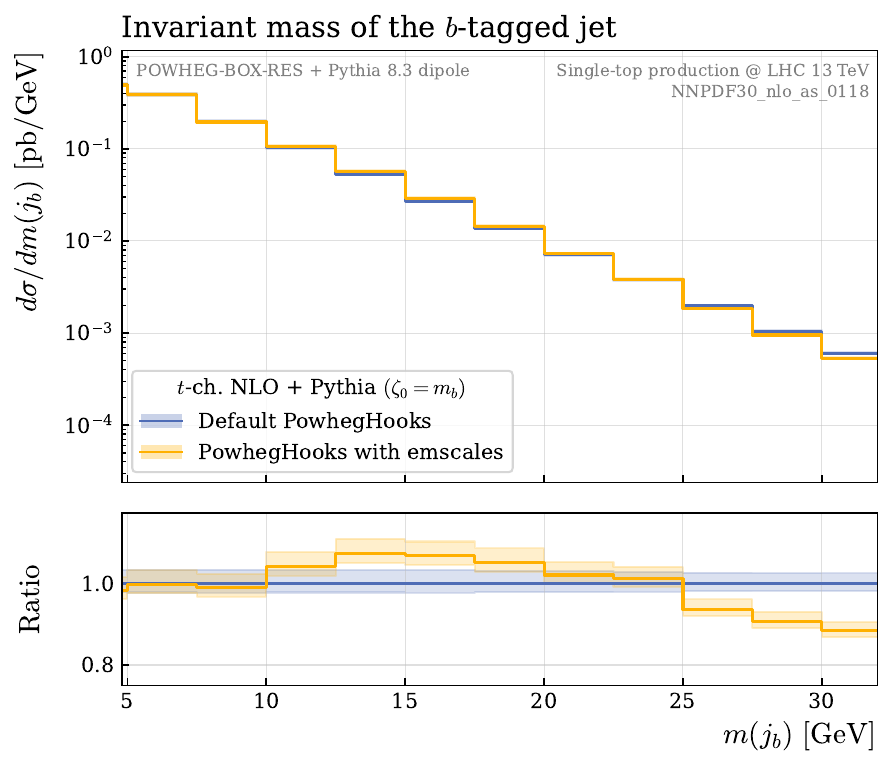}
\caption{}
\end{subfigure}

\vspace{0.5cm}

\begin{subfigure}[b]{.45\linewidth}
\includegraphics[width=1\textwidth,height=6cm,keepaspectratio]{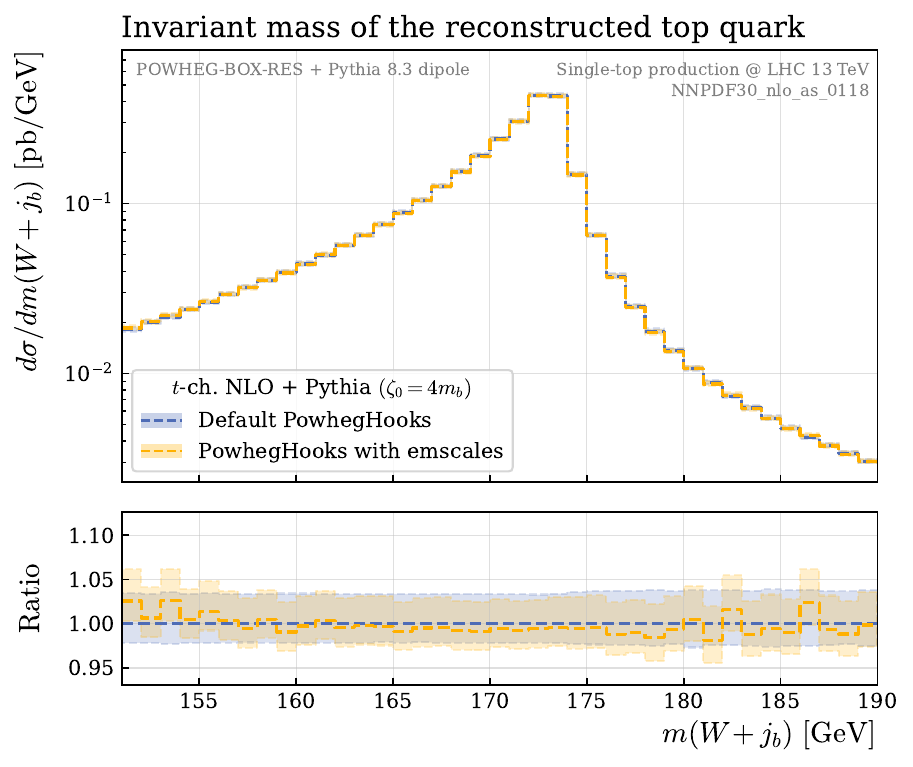}
\caption{}
\end{subfigure}
\hfill
\begin{subfigure}[b]{.45\linewidth}
\includegraphics[width=1\textwidth,height=6cm,keepaspectratio]{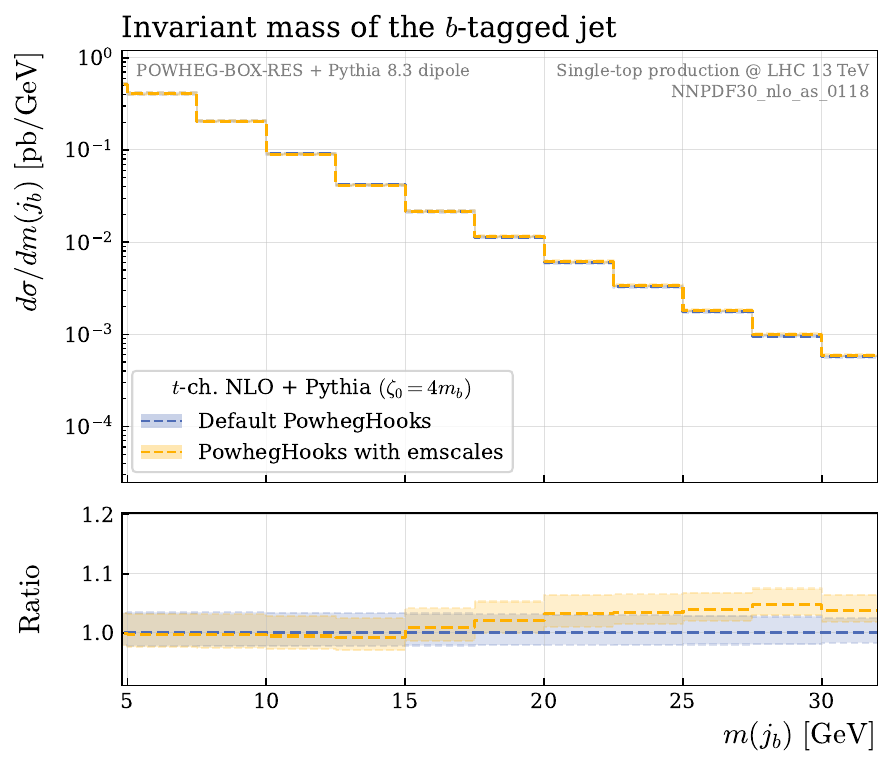}
\caption{}
\end{subfigure}

\caption{Effect of the choice of \texttt{UserHook} on the invariant mass of the reconstructed top quark (left) and of the $b$-tagged jet (right). Blue corresponds to the default choice described in Sec.~\ref{sec:matching_py8}, while yellow corresponds to the choice described in App.~\ref{sec:hooks}. The upper two figures show differences between the two implementations for $\zeta_0=m_b$, while these differences disappear for the larger value $\zeta_0=4m_b$. Error bands represent the seven-point scale variation uncertainty. We stress that we do not recommend the larger $\zeta_0$ value for phenomenological applications. }
\label{fig:diffhooks}
\end{figure}

The numerical impact of replacing the reconstructed veto scales by the stored \powheg{} scales is shown for $t$-channel in Fig.~\ref{fig:diffhooks}. 
We compare the default \texttt{PowhegHooks} implementation, shown in blue, with the \texttt{emscales}-based prescription, shown in yellow, for two different choices of the shower cutoff. 
The left panels show the invariant mass of the reconstructed top quark, while the right panels show the invariant mass of the $b$-tagged jet. 
For the lower value, $\zeta_0=m_b$, shown in the upper panels, the two prescriptions lead to modest but visible differences, reaching the level of about $10\%$ in some regions of the distributions. 
These differences are strongly reduced when the cutoff is raised to $\zeta_0=4m_b$, shown in the lower panels, where the two predictions are largely compatible within the scale-variation uncertainties. 
This behaviour indicates that the numerical sensitivity to the precise definition of the resonance-decay veto scale is concentrated in the region in which the shower is allowed to evolve further into the low-scale regime.
For $s$-channel, we observe a very small difference already at the $\zeta_0=m_b$ scale due to the high fraction of regular events, which do not include multiple emissions.
If neither of the prescriptions is clearly preferred, the difference between the two may be considered as an additional matching uncertainty.

We do not implement the alternative \texttt{scaleResonanceVeto} approach available in the \texttt{bb4l} interface. 
Although this prescription is equivalent to the \texttt{doVetoFSREmission} one at leading-logarithmic accuracy, the two can differ in the hard-emission region.
The \texttt{scaleResonanceVeto} hook sets the starting scale of the resonance-decay shower, after which the evolution is governed by \pythia's own ordering variable. 
This variable agrees with the \powheg{} hardness only in the soft-collinear limit. 
By contrast, \texttt{doVetoFSREmission} provides access to each generated branching and allows its hardness to be evaluated explicitly using the same definition as for the stored \powheg{} emission.

\newpage

\bibliographystyle{JHEP.bst}
\bibliography{References}

\end{document}